\documentclass[preprint,showkeys]{revtex4}

\usepackage{amssymb}
\usepackage{graphicx}
\newcommand{\be}{\begin{eqnarray}}
\newcommand{\ee}{\end{eqnarray}}

\begin{document}

\title{ Transverse charge and magnetization densities in holographic QCD}
\author{\bf Dipankar Chakrabarti and Chandan Mondal}
\affiliation{Department of Physics, Indian Institute of Technology Kanpur, Kanpur-208016, India.}
%
%

\begin{abstract}
We present a  study of flavor structures of  the transverse charge and anomalous magnetization densities for both unpolarized and transversely polarized nucleons. We consider two different models for the  electromagnetic form factors in holographic QCD. The flavor form factors are obtained by decomposing the Dirac and Pauli form factors for nucleons using the charge and isospin symmetry. The results are compared with two standard phenomenological parametrizations.
\end{abstract}
\maketitle



\section{ Introduction}

Form factors of the nucleons  provide us with crucial information about the internal structure of the nucleons  and have been measured in  many experiments. The charge and magnetization densities in the transverse plane are defined as  the Fourier transforms of the electromagnetic form factors.   The transverse densities  are also intimately related  to  the generalized parton distributions (GPDs) with zero skewness.  The contributions of individual quark  to the nucleon charge and magnetization densities are obtained  from the  flavor decompositions of the transverse densities.  The transverse densities corresponding to individual quarks are   given by the Fourier transforms of the GPDs in the transverse coordinate or impact parameter space\cite{burk}. 
 The form factor involves initial and final  states with different momenta  and  three dimensional Fourier transforms  cannot be interpreted as  densities whereas the transverse densities defined at fixed light front time are free from this difficulty and  have proper density interpretation \cite{miller09,miller10,venkat}.
 
Recently, AdS/QCD has emerged as one of the most promising techniques to unravel the structure of mesons and nucleons.  The AdS/CFT conjecture\cite{maldacena}   relates a gravity theory in $AdS_{d+1}$ to a conformal theory   at the  $d$ dimensional boundary.   There are many applications of AdS/CFT to investigate the QCD phenomena\cite{PS,costa1}.  A boundary condition  in the fifth dimension $z$ in $AdS_5$  breaks the conformal invariance and allows QCD mass scale and confinement. In  the hard-wall model, an IR cutoff is set at  $z_0=1/\Lambda_{QCD}$ while in soft-wall model, a confining potential in $z$ is introduced. 
There is an exact correspondence between the holographic variable $z$ and the light front transverse variable $\zeta$ which measures the separation of the quark and  gluonic constituents in the hadron\cite{BT00,BT01}.
 The AdS/QCD for the baryon  has been developed by  several groups 
\cite{BT00,BT01,SS,katz,ads1,ads2,AC}.
Though it gives only a semiclassical approximation of QCD, so far this method has  been successfully applied to  describe many hadron properties e.g., hadron mass spectrum, parton distribution functions, GPDs, meson and nucleon form factors,  structure functions etc\cite{BT1,BT1b,AC,BT2,HSS,vega,AC4,CM,CM2}.  AdS/QCD wave functions  are used to predict the experimental data for $\rho$ meson electroproduction \cite{forshaw}. 
AdS/QCD has also been successfully applied in the meson sector to predict the branching ratio for decays of $\bar{B^0}$ and $\bar{B_s^0}$ into $\rho$ mesons \cite{ahmady1}, isospin asymmetry and branching ratio for the $B\to K^*\gamma$ decays \cite{ahmady2}, transition form factors\cite{AC2,ahmady3}, etc.  There are many other applications in the baryon sector e.g.,  semi-empirical hadronic momentum density distributions in the transverse plane   have been calculated in\cite{AC3},  in \cite{hong2}, the form facfor of spin $3/2$ baryons ($\Delta$ resonance) and also the transition form factor between $\Delta$ and nucleon have been studied, an AdS/QCD   model has been proposed to study the baryon spectrum at finite temperature\cite{li} etc.

 The flavor decompositions of the nucleon form factors in a light-front quark model with SU(6) spin-flavor symmetry have been  studied in detail in \cite{CM2} and shown to agree with experimental data. It is interesting and instructive to study the transverse densities and their flavor decomposition in holographic QCD. There are two different holographic QCD  models for nucleon form factors developed by Abidin and Carlson\cite{AC} and Brodsky and Teramond\cite{BT2}.  Here, we present a detailed analysis of the transverse densities in  the two models.

Model-independent transverse charge densities for nucleons have been studied in \cite{miller07} whereas the charge densities in the transverse plane for a transversely polarized nucleon are shown in \cite{vande,selyugin}.  In \cite{MVT}, the long range behavior of the unpolarized quark 
transverse charge density of the nucleons has been studied.   Transverse charge and magnetization densities in the nucleon's  chiral periphery(i.e., at a distance $b={\cal{O}}(1/m_\pi)$) using methods of dispersion analysis and chiral effective field theory have been analysed in \cite{weiss}. The transverse densities for the quarks are studied in a chiral quark-soliton model in \cite{silva}.
Using Laguerre-Gaussian expansion, Kelly \cite{kelly02} proposed a parametrization of  the nucleon Sachs form factors in terms of  charge and magnetization densities. A study of flavor dependence of the transverse densities in a GPD model has been reported in \cite{liuti}.

 In \cite{vega}, the  nucleon transverse charge and magnetization densities have been evaluated in the model developed in \cite{AC}. In this work, we show   the flavor decompositions of the  transverse densities of the nucleons in two different models  in the framework of AdS/QCD and compare with the two global parametrizations  of Kelly \cite{kelly04} and Bradford $at~ el$ \cite{brad}. By decomposing  the nucleon form factors $F_1$ and $F_2$ using the charge and isospin symmetry, we obtain the flavor form factors $F_1^q$ and $F_2^q$  for the quarks. The Fourier transforms of these electromagnetic form factors give the charge and magnetization densities in the transverse plane.

 The paper is organized as follows. A brief description of the   form factors in AdS/QCD  has been has given in Sect.\ref{ads}. In Sect.\ref{density}, the charge and magnetization densities for both unpolarized and transversely polarized nucleons have been studied. The individual flavor contributions are also studied in this section. Then we provide a brief summary in Sect.\ref{concl}.

\vskip0.2in
\noindent
\section{Nucleon and flavor form factors in AdS/QCD}\label{ads}
Here we consider the soft wall model of AdS/QCD, where in place of a sharp cutoff in  z, one introduces a potential.
 The  action in soft model  is written as\cite{BT2}
 \be
S&=&\int d^4x dz \sqrt{g}\Big( \frac{i}{2}\bar\Psi e^M_A\Gamma^AD_M\Psi -\frac{i}{2}(D_M\bar{\Psi})e^M_A\Gamma^A\Psi\nonumber\\
&&-\mu\bar{\Psi}\Psi-V(z)\bar{\Psi}\Psi\Big),\label{action}
\ee
where $e^M_A=(z/R)\delta^M_A$ is the inverse  vielbein and $V(z)$ is the confining potential which breaks the conformal invariance and 
 $R$ is the AdS radius. The covariant derivative  is $D_M=\partial_M-\frac{i}{2}\omega_M^{AB}\Sigma_{AB}$ where $\omega_M^{AB}=(\eta^{Az}\delta_M^B-\eta^{Bz}\delta_M^A)/z$ and $\Sigma_{AB}=\frac{i}{4}[\Gamma_A,\Gamma_B]$.
  
 The   Dirac equation in AdS derived from the above action is given by
\be
i\Big(z \eta^{MN}\Gamma_M\partial_N+\frac{d}{2}\Gamma_z\Big)\Psi -\mu R\Psi-RV(z)\Psi=0.\label{ads_DE}
\ee
With $z$ identified as the  light front transverse impact variable $\zeta$ which gives the separation of the  quark and 
gluonic constituents in the hadron, it is possible to extract  the light-front wave functions for the hadron. 
In $d=4$ dimensions, $\Gamma_A=\{\gamma_\mu, -i\gamma_5\}$.
To map with the light front wave equation, we identify $z\to\zeta$, 
where $\zeta$
is the light front transverse variable,  and substitute $\Psi(x,\zeta)=e^{-iP\cdot x}\zeta^2\psi(\zeta)u(P)$ in Eq.(\ref{ads_DE})  and set $\mid \mu R\mid=\nu+1/2$ where  $\nu$ is related with the orbital angular momentum by $\nu=L+1$ .
 For linear confining potential  $U(\zeta)=(R/\zeta)V(\zeta)=\kappa^2\zeta$, we get the light front wave equation for the baryon in $2\times 2$ spinor representation as
\be
 \big(-\frac{d^2}{d\zeta^2}-\frac{1-4\nu^2}{4\zeta^2}+\kappa^4\zeta^2&+&2(\nu+1)\kappa^2\Big)\psi_+(\zeta)\nonumber\\
 &=&{\cal{M}}^2\psi_+(\zeta),\\
 \big(-\frac{d^2}{d\zeta^2}-\frac{1-4(\nu+1)^2}{4\zeta^2}&+&\kappa^4\zeta^2+2\nu\kappa^2\Big)\psi_-(\zeta)\nonumber\\
 &=&{\cal{M}}^2\psi_-(\zeta).
 \ee 
  In case of mesons, the similar potential $\kappa^4\zeta^2$ appears in the Klein-Gordon equation  which can be generated by introducing a dilaton background $\phi=e^{\pm\kappa^2 z^2}$ in the AdS space which breaks the conformal invariance. But in case of baryon, the dilaton can be scaled out by a field redefinition\cite{BT2}. So, the confining potential for baryons cannot be produced by dilaton and is put in by hand in the soft wall model. The form of the confining potential $(\kappa^4\zeta^2)$ is unique for both the meson and baryon sectors \cite{Dosch}.
  The solutions of the above equations are 
  \be
  \psi_+(z) &\sim & \zeta^{\nu+1/2} e^{-\kappa^2 \zeta^2/2}L_n^\nu(\kappa^2\zeta^2)\label{psi+},\\
\psi_-(z) &\sim & \zeta^{\nu+3/2} e^{-\kappa^2 \zeta^2/2}L_n^{\nu+1}(\kappa^2\zeta^2)\label{psi-}.
\ee

\begin{figure}[htbp]
\begin{minipage}[c]{0.98\textwidth}
\small{(a)}
\includegraphics[width=7cm,height=6cm,clip]{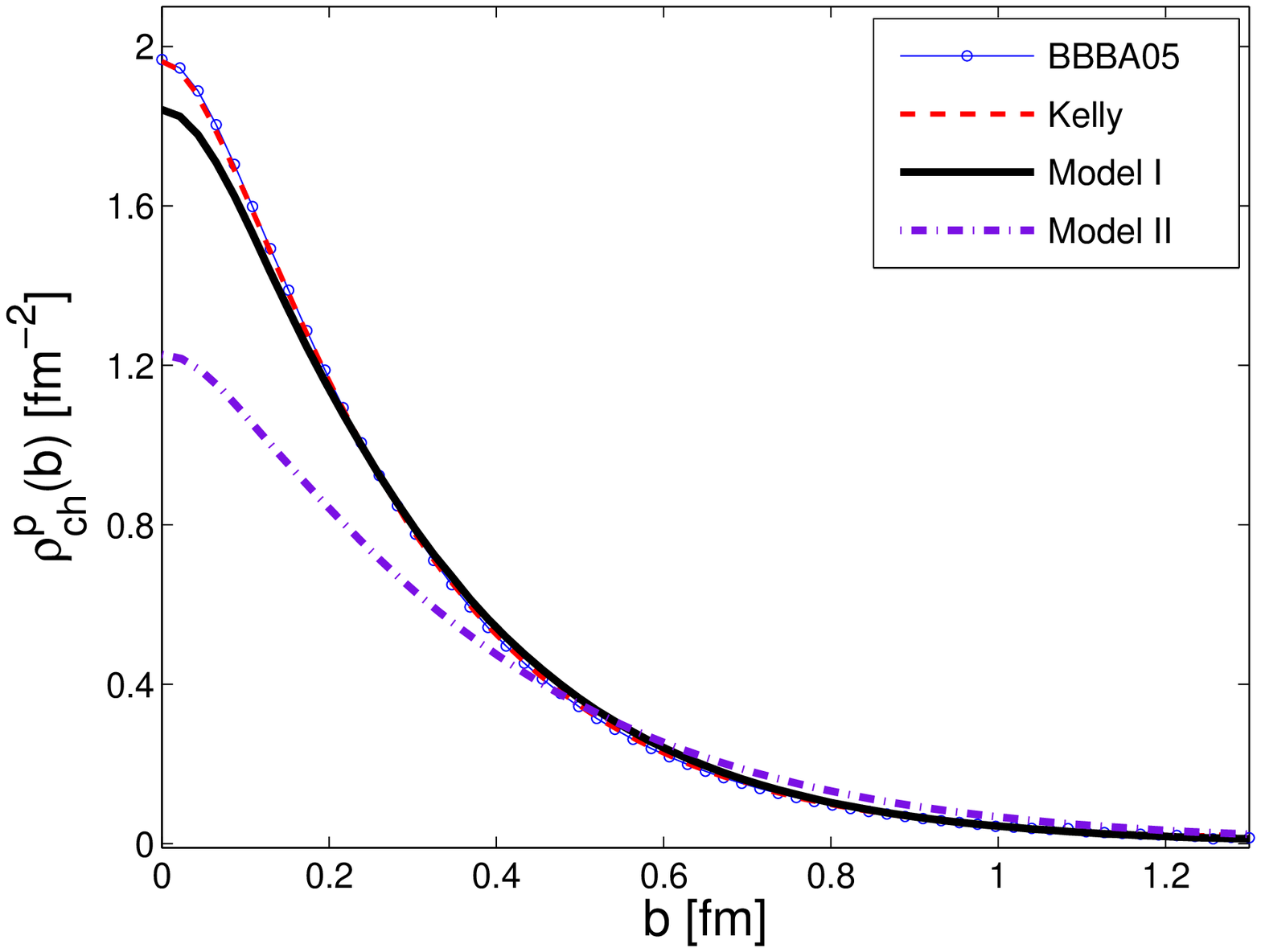}
\hspace{0.1cm}%
\small{(b)}\includegraphics[width=7cm,height=6cm,clip]{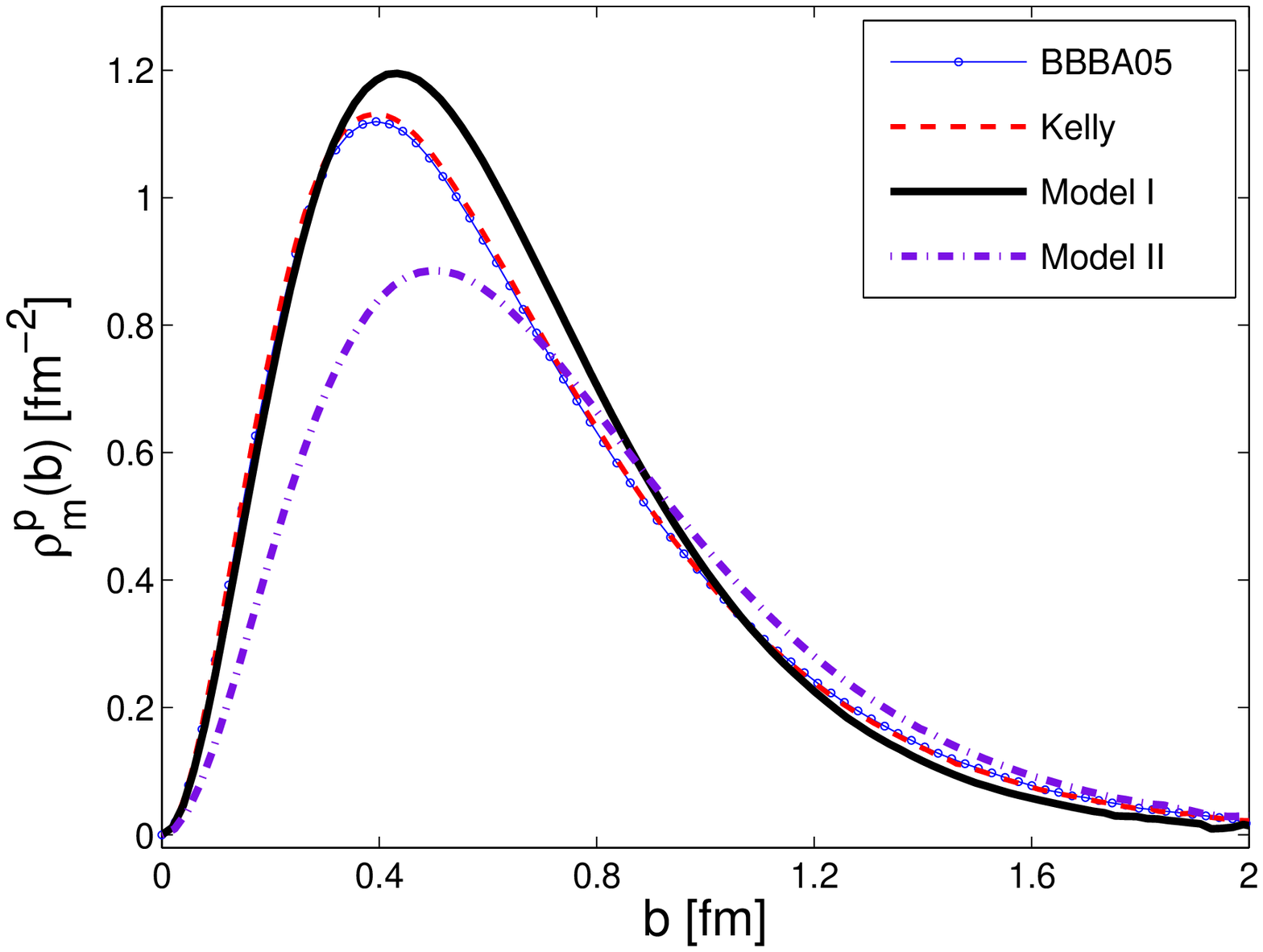}
\end{minipage}
\begin{minipage}[c]{0.98\textwidth}
\small{(c)}\includegraphics[width=7cm,height=6cm,clip]{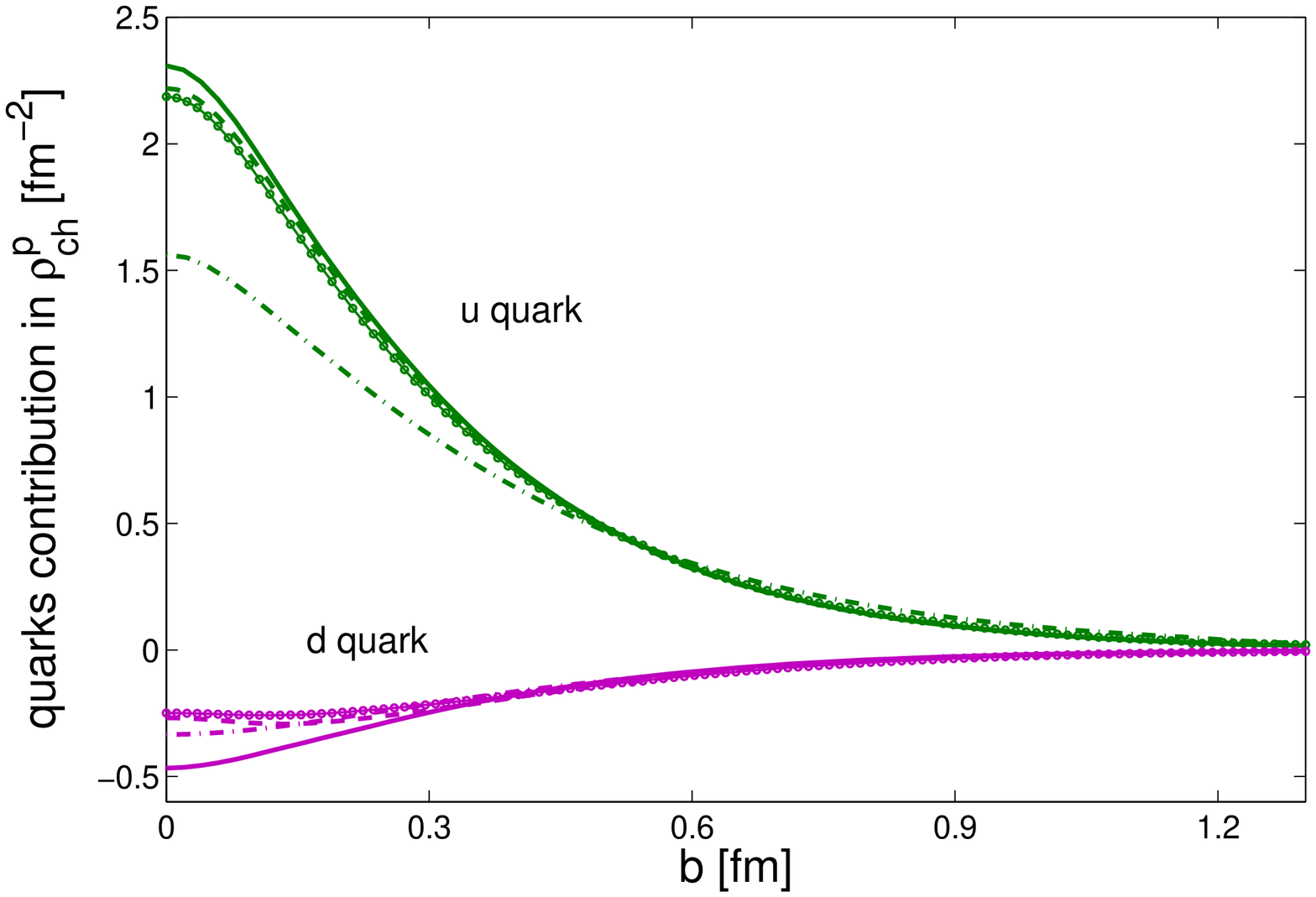}
\hspace{0.1cm}%
\small{(d)}\includegraphics[width=7cm,height=6cm,clip]{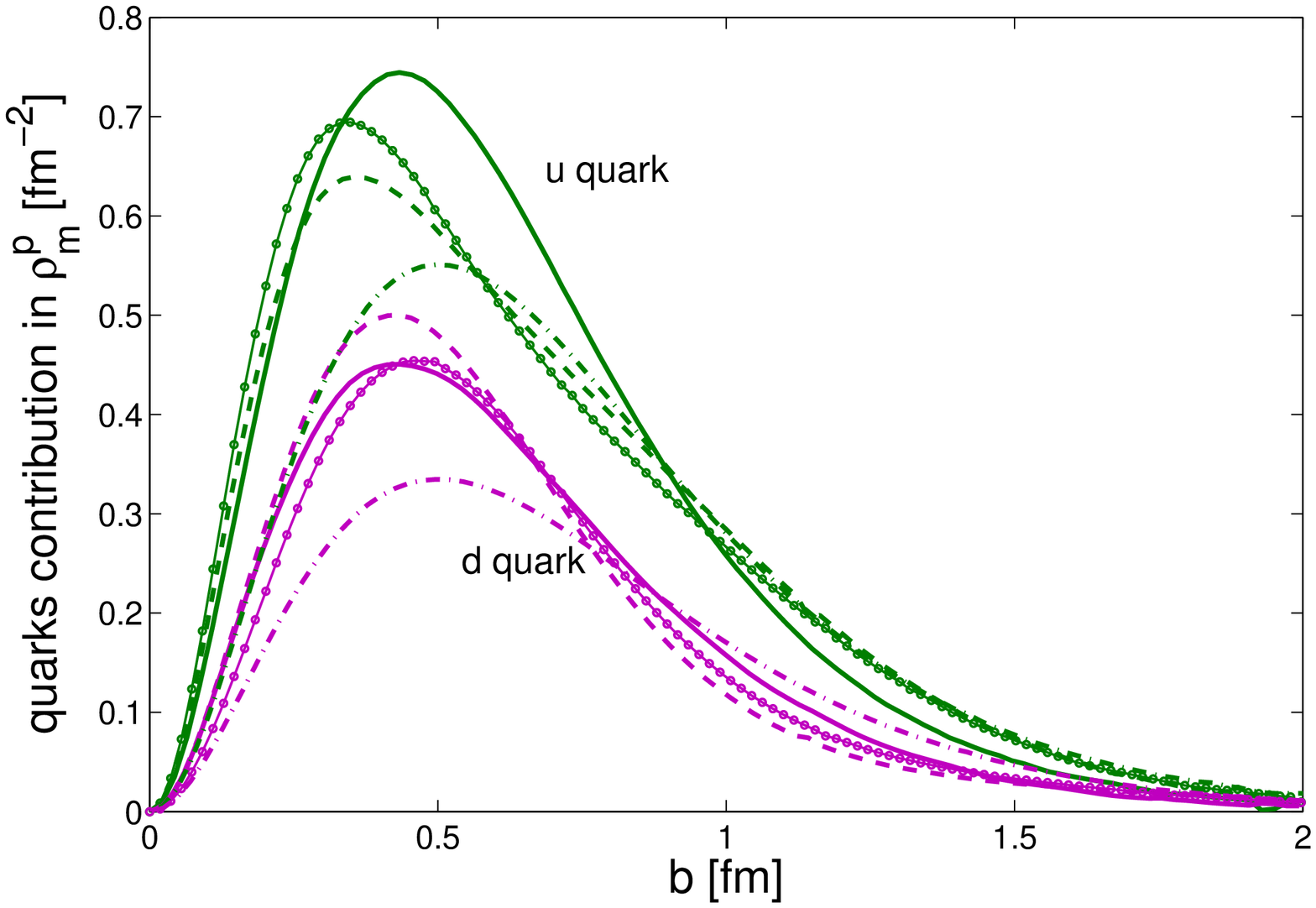}
\end{minipage}
\caption{\label{proton_flavors}(Color online) Plots of  flavor dependent transverse charge and anomalous magnetization densities for proton. (a) and (b) represent $\rho_{ch}$ and ${\rho}_m$ for the proton. (c) and (d) represent the contributions from different flavors.  Dashed line represents the parametrization of  Kelly \cite{kelly04},  and  the line with circles represents the parametrization of  Bradford $at~el$ \cite{brad}; 
 the solid line is for Model-I and dot-dashed line is for Model-II. In (c) and (d) $u$ and $d$ quark contributions are plotted in different colors.
}
\end{figure}
\begin{figure}[htbp]
\begin{minipage}[c]{0.98\textwidth}
\small{(a)}
\includegraphics[width=7cm,height=6cm,clip]{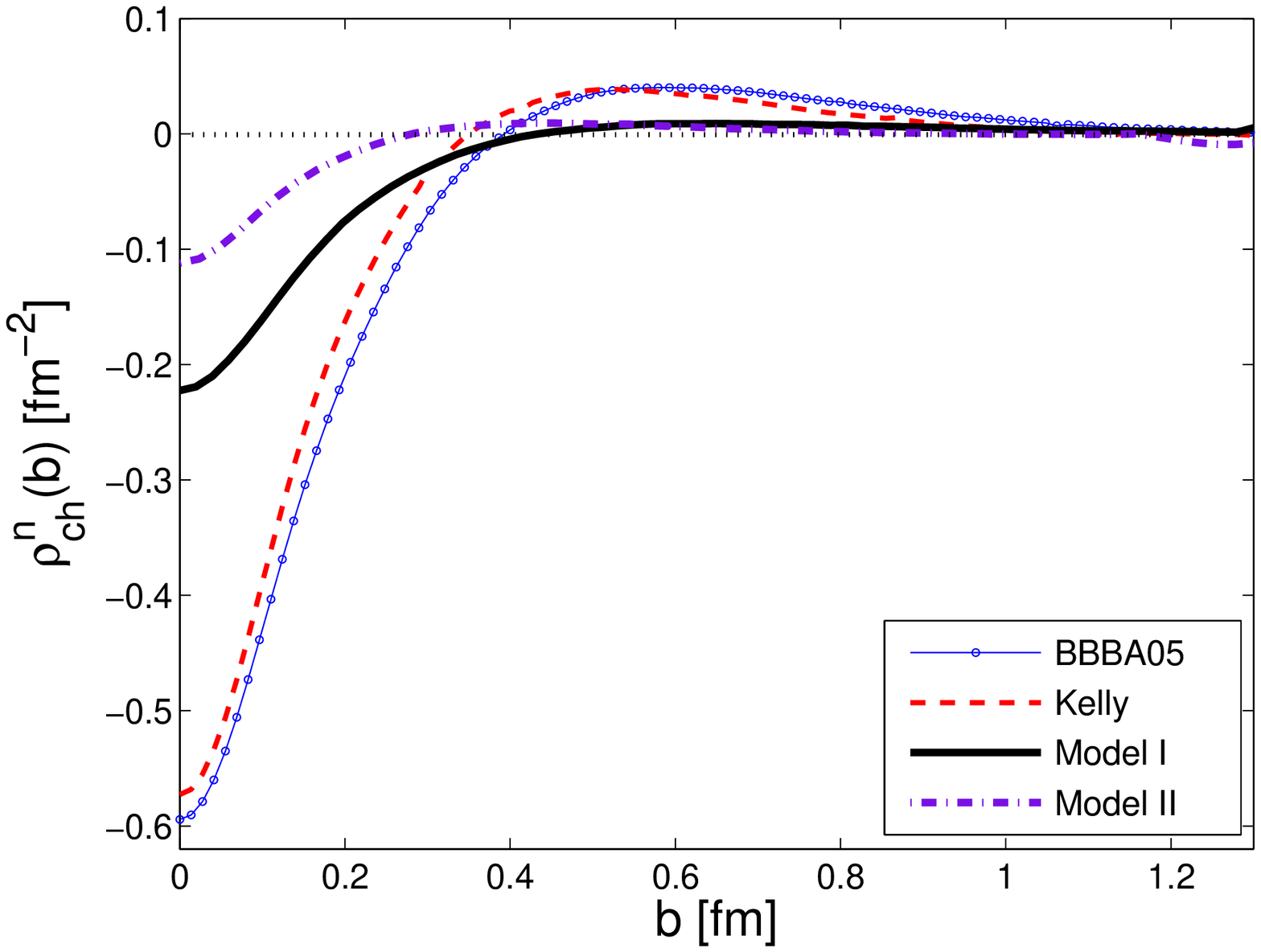}
\hspace{0.1cm}%
\small{(b)}\includegraphics[width=7cm,height=6cm,clip]{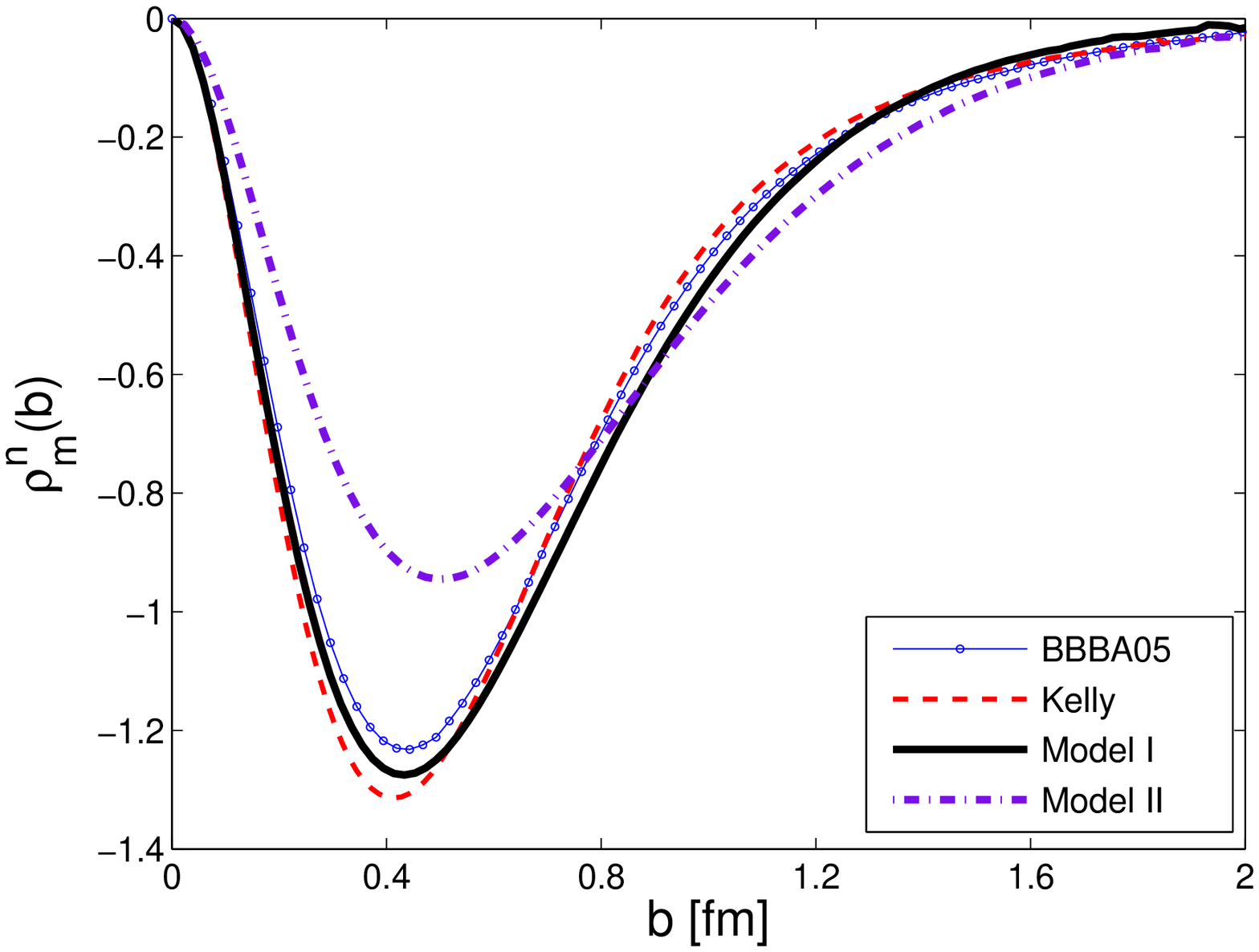}
\end{minipage}
\begin{minipage}[c]{0.98\textwidth}
\small{(c)}\includegraphics[width=7cm,height=6cm,clip]{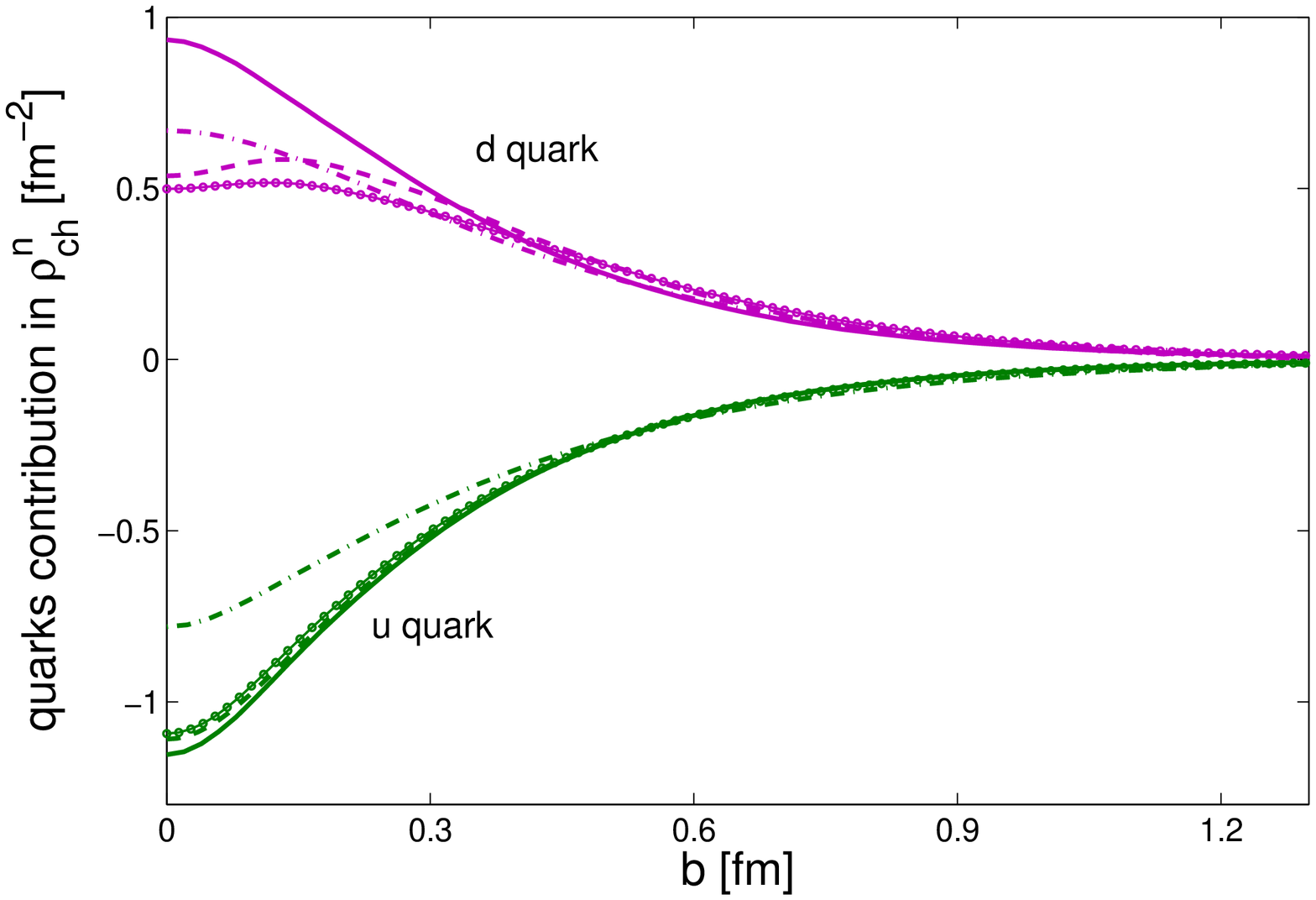}
\hspace{0.1cm}%
\small{(d)}\includegraphics[width=7cm,height=6cm,clip]{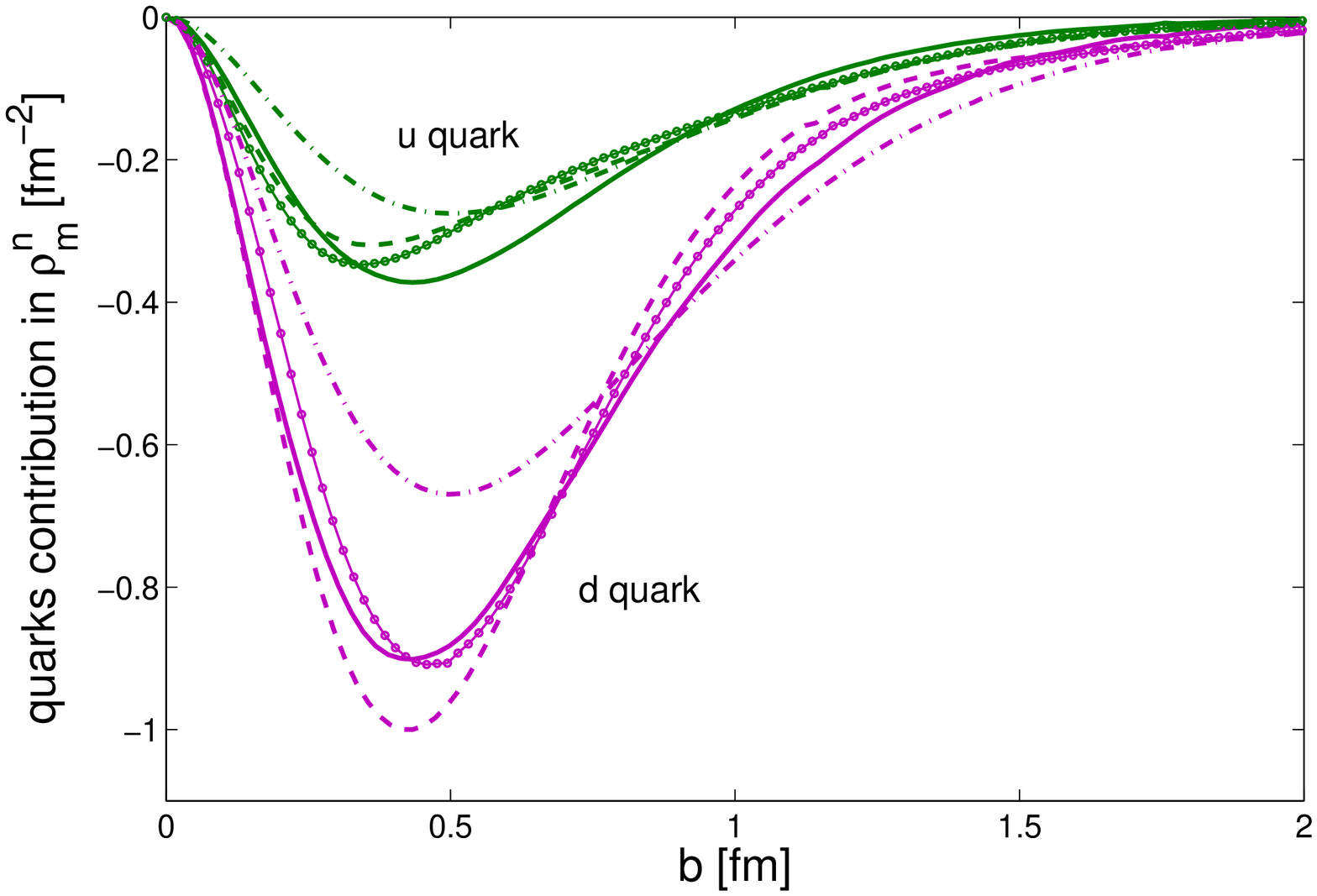}
\end{minipage}
\caption{\label{neutron_flavors}(Color online) Plots of  flavor dependent transverse charge and anomalous magnetization densities for neutron. (a) and (b) represent $\rho_{ch}$ and ${\rho}_m$ for the neutron. (c) and (d) represent the contributions from different flavors. Dashed line represents the parametrization of  Kelly \cite{kelly04},  and the  line with circles represents the parametrization of  Bradford $at~el$ \cite{brad}; 
 the solid line is for Model-I and dot-dashed line is for Model-II. In (c) and (d) $u$ and $d$ quark contributions are plotted in different colors.
}
\end{figure}

{\bf Model-I}

By Model-I, we refer to the AdS/QCD model for form factors proposed by Brodsky and Teramond \cite{BT2}.
The SU(6) spin-flavor symmetric quark model is constructed in the AdS/QCD by weighing the different Fock-state components by the charges and spin projections of the partons as dictated by the symmetry.   In the light-front quark model the Dirac and Pauli form factors are described by the spin-nonflip and spin-flip matrix elements of the electromagnetic current $J^+=e_q\bar\psi\gamma^+\psi$ \cite{BD}.



The Dirac form factors for the nucleons are  obtained in this model  as
\be
F_1^p(Q^2)&=&R^4\int \frac{dz}{z^4} V(Q^2,z)\psi^2_+(z),\label{F1p}\\
F_1^n(Q^2) &=& -\frac{1}{3}R^4\int \frac{dz}{z^4} V(q^2,z)(\psi^2_+(z)-\psi^2_-(z)),\label{F1n}
\ee
and the Pauli form factor is written as 
\be
F_2^{p/n}(Q^2) &\sim &  \int \frac{dz}{z^3} \psi_+(z)V(Q^2,z) \psi_-(z).\label{F2}
\ee

The form factors are  normalized to $F_1^p(0)=1, F_1^n(0)=0 $ and $F_2^{p/n}(0)=\kappa_{p/n}$, where $\kappa_{p/n} $ is the anomalous magnetic moments for the proton/neutron. Note that the Pauli form factor is not mapped properly in this model.  In the light front quark model, it is defined as the spin flip matrix element of $J^+$ current but the AdS action cannot produce this term and the Pauli form factor is   put in for phenomenological purposes. The twist-3 nucleon wave functions in the soft wall model are  
\be
\psi_+(z) &= &\frac{\sqrt{2}\kappa^2}{R^2}z^{7/2} e^{-\kappa^2 z^2/2}\label{psi+},\\
\psi_-(z) &= & \frac{\kappa^3}{R^2}z^{9/2} e^{-\kappa^2 z^2/2}\label{psi-}.
\ee
The bulk-to-boundary propagator 
is given by
\be
 V(Q^2,z)=\Gamma(1+\frac{Q^2}{4\kappa^2})U\big(\frac{Q^2}{4\kappa^2},0,\kappa^2 z^2),\label{propagator}
 \ee
 where $U(a,b,z)$ is the Tricomi confluent hypergeometric function.
The bulk-to-boundary propagator, Eq. (\ref{propagator}), can be written in a simple integral form \cite{Rad,BT2}
\be
\! V(Q^2,z)=\kappa^2z^2\int_0^1\!\frac{dx}{(1-x)^2} x^{Q^2/(4\kappa^2)} e^{-\kappa^2 z^2 x/(1-x)}.
 \ee
 We refer to  the formulas for the form factors given in Eqs.(\ref{F1p},\ref{F1n} and \ref{F2}) as Model-I. 
  It has been shown\cite{CM,CM2} that the form factors for the nucleons agree with experimental data for $\kappa=0.4066~ GeV$.

 {\bf Model-II}
 
 The other model of the form factors was formulated by Abidin and Carlson\cite{AC}. Since the action defined in Eq.(\ref{action}) cannot produce the spin flip (Pauli) form factors, they introduced an additional gauge invariant non-minimal coupling. This additional term also gives an anomalous contribution to the Dirac form factor. In this model the form factors are given by\cite{AC}
 \be
 F_1^p(Q^2) &=& C_1(Q^2)+\eta_p C_2(Q^2),\label{F1pM2}\\
 F_1^n(Q^2)&=& \eta_n C_2(Q^2),\label{F1nM2}\\
 F_2^p(Q^2)&=& \eta_p C_3(Q^2),\label{F2pM2}\\
 F_2^n(Q^2)&=& \eta_n C_3(Q^2).\label{F2nM2}
 \ee
 where
  \be
 C_1(Q^2) &=& \frac{a+6}{(a+1)(a+2)(a+3)},\\
 C_2(Q^2) &=& \frac{2a(2a-1)}{(a+1)(a+2)(a+3)(a+4)},\\
 C_3(Q^2) &=& \frac{48}{(a+1)(a+2)(a+3)},
 \ee
 where $a=Q^2/(4\kappa^2)$. The value of  $\kappa$  is fixed by simultaneous fit to proton and rho meson mass and the best fit gives the value
  $\kappa=0.350 GeV$.  The other parameters are determined from the normalization conditions of the Pauli form factor at $Q^2=0$ and are given by $\eta_p=0.224$ and $\eta_n=-0.239$ \cite{AC}. We refer the form factors given by Eqs. (\ref{F1pM2}-\ref{F2nM2})  as Model-II.
  
The Pauli form factors in these two models are identical,  the main difference  is in the Dirac form factor.  
In model-II, there is an additional contribution to the Dirac form factor from the non-minimal coupling term.
It should be mentioned here that the Pauli form factors in  the AdS/QCD models are mainly of  phenomenological origin.
Since the action in Eq.(\ref{action}) cannot produce the spin flip term,  in Model-II, a non-minimal coupling term is added to generate the Pauli form factors. This additional term gives contribution to the Dirac form factors also. The major difference between these two models is that  in the Model-I, 
  the anomalous contributions to the Dirac form factors are not considered. 
The additional contribution from the nonminimal coupling to the Dirac form factor  corresponds to higher twist  and not included in model-I, while they are included  in the model-II.  
 
 Under  the charge  and isospin symmetry it is straightforward to write down the flavor decompositions of the nucleon form factors    as
\be
F_i^u=2F_i^p+F_i^n ~~{\rm and} ~~F_i^d=F_i^p+2F_i^n,~~(i=1,2)
\ee
with the normalizations $F_1^u(0)=2, F_2^u(0)=\kappa_u$ and $F_1^d(0)=1, F_2^d(0)=\kappa_d$ where the anomalous magnetic moments for the up and down quarks are $\kappa_u=2\kappa_p+\kappa_n=1.673$ and $\kappa_d=\kappa_p+2\kappa_n=-2.033$.
 It was shown in \cite{Cates} that though the ratio of Pauli and Dirac form factors for the proton $F_2^p/F_1^p \propto 1/Q^2$,  the $Q^2$ dependence is almost constant for the ratio of the quark form factors $F_2/F_1$ for both $u$ and $d$.  
 \begin{figure}[htbp]
\begin{minipage}[c]{0.98\textwidth}
\small{(a)}
\includegraphics[width=7cm,height=6cm,clip]{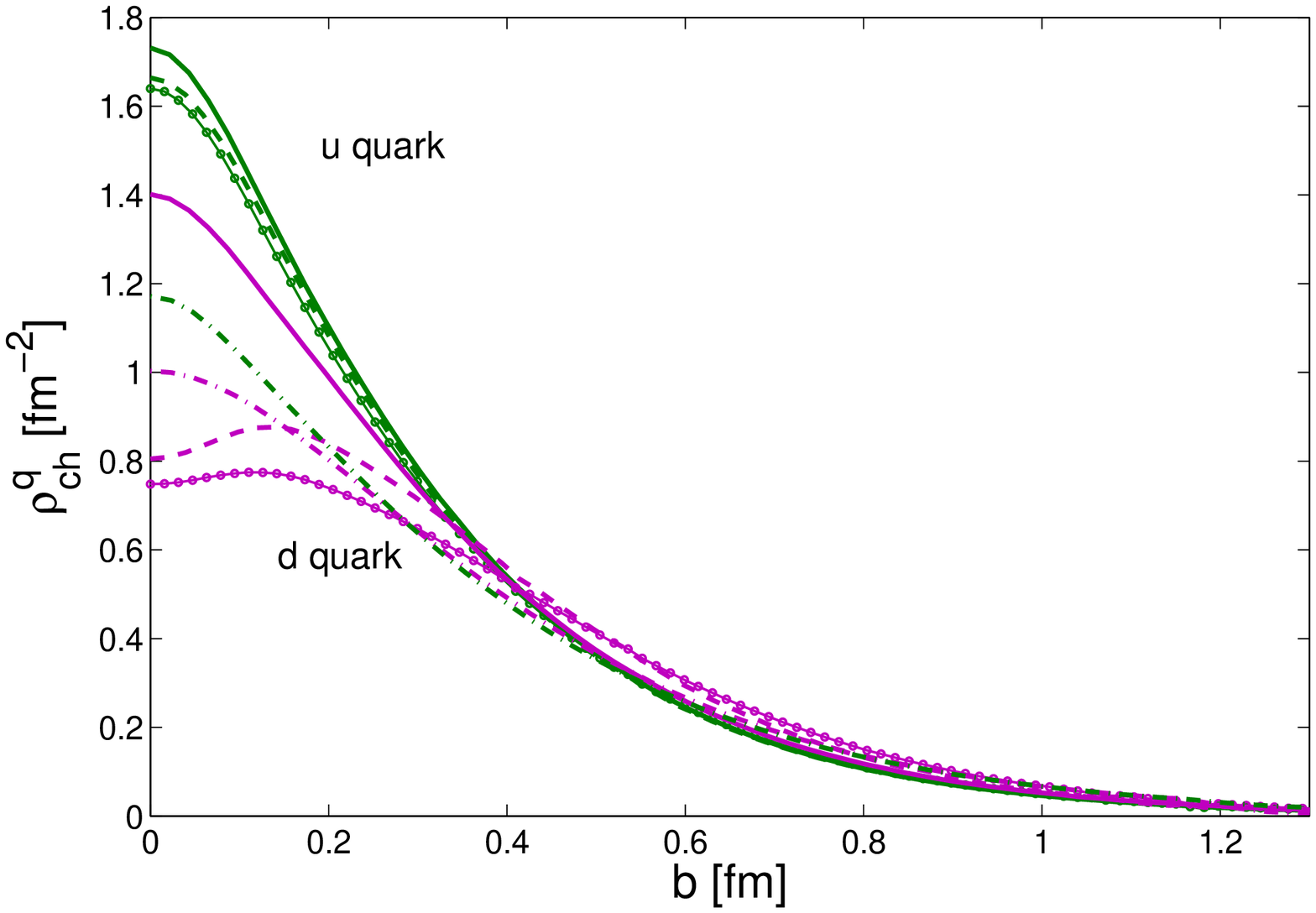}
\hspace{0.1cm}%
\small{(b)}\includegraphics[width=7cm,height=6cm,clip]{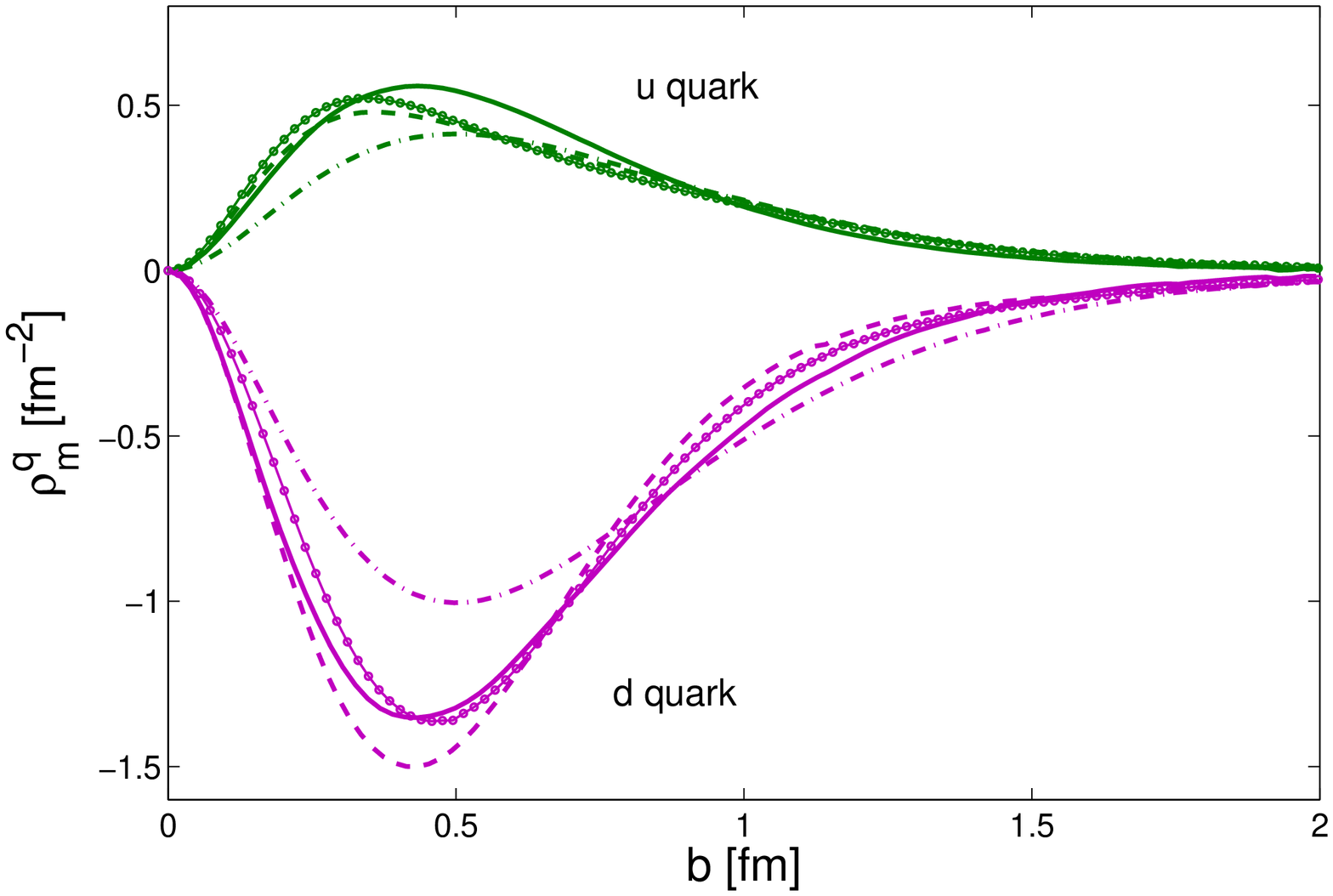}
\end{minipage}
\caption{\label{flavors}(Color online) Plots of quarks transverse charge and anomalous magnetization densities. (a) represent $\rho_{ch}^q$ and (b) represent ${\rho}_m^q$. 
Dashed line represents the parametrization of  Kelly \cite{kelly04},  and  line with circles represents the parametrization of  Bradford $at~el$ \cite{brad}; 
 the solid line is for Model-I and dot-dashed line is for Model-II.  Densities for $u$ and $d$ quark  are plotted in different colors.
 }
\end{figure}
\begin{figure}[htbp]
\begin{minipage}[c]{0.98\textwidth}
\small{(a)}
\includegraphics[width=7cm,height=6cm,clip]{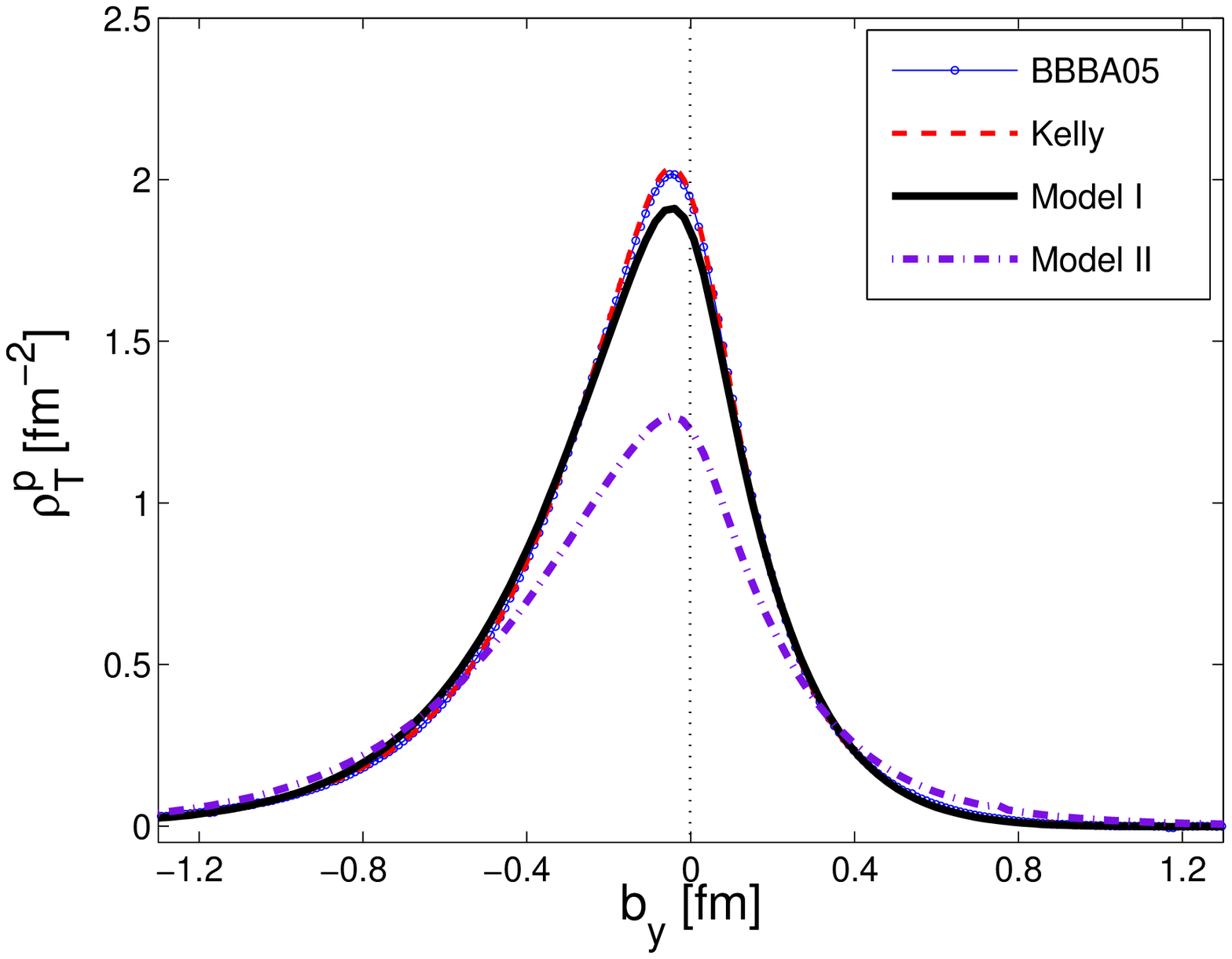}
\hspace{0.1cm}%
\small{(b)}\includegraphics[width=7cm,height=6cm,clip]{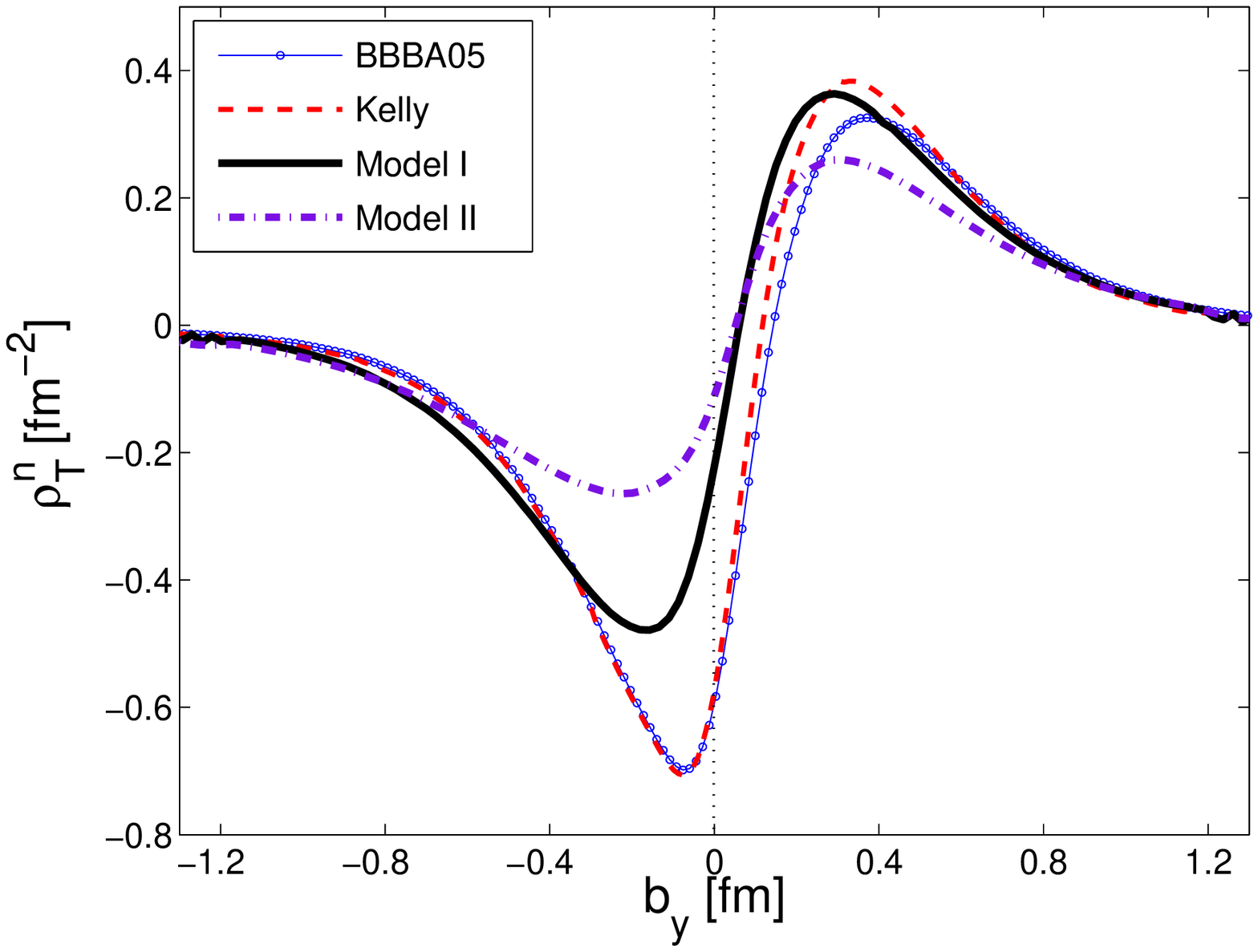}
\end{minipage}
\begin{minipage}[c]{0.98\textwidth}
\small{(c)}
\includegraphics[width=7cm,height=6cm,clip]{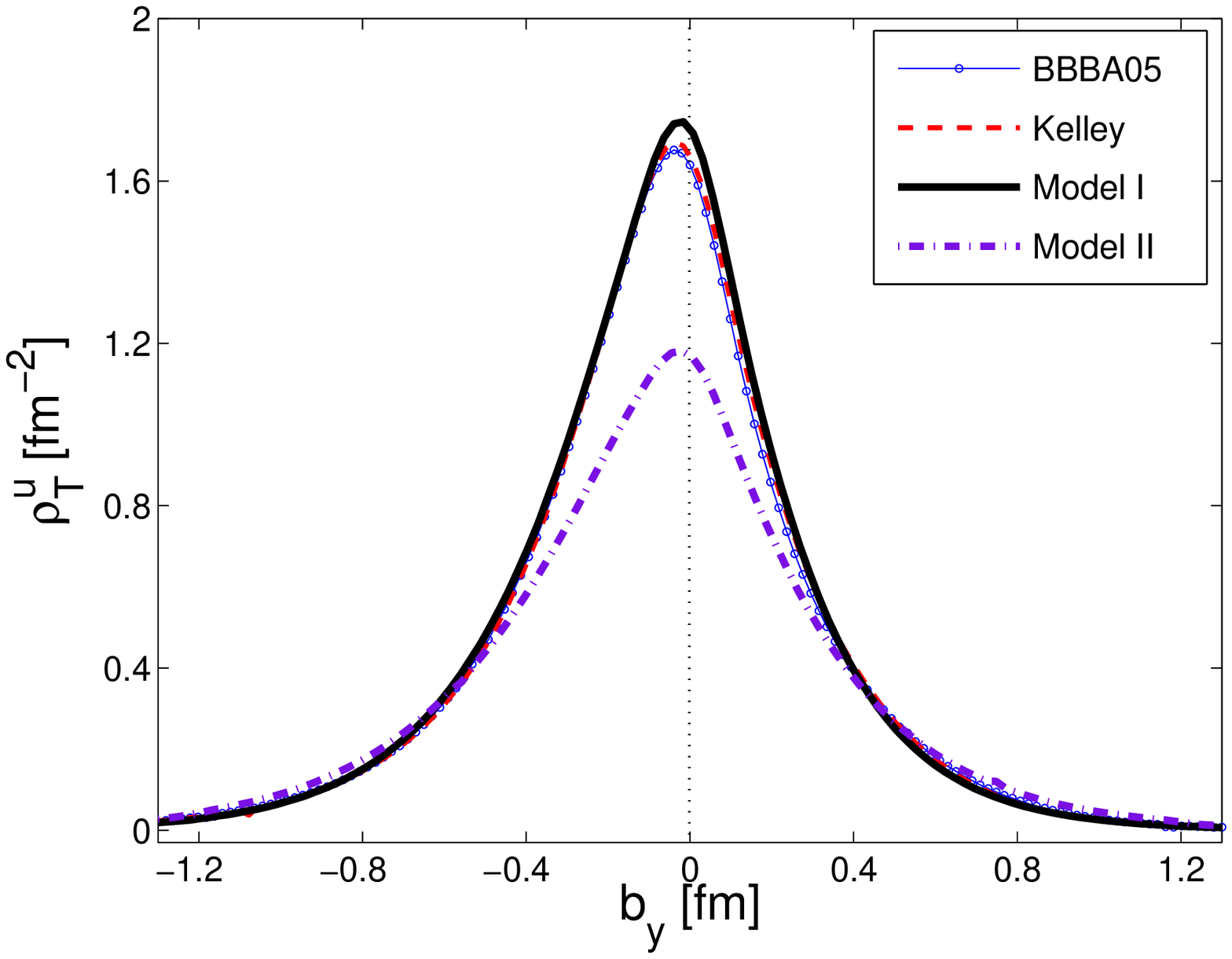}
\hspace{0.1cm}%
\small{(d)}\includegraphics[width=7cm,height=6cm,clip]{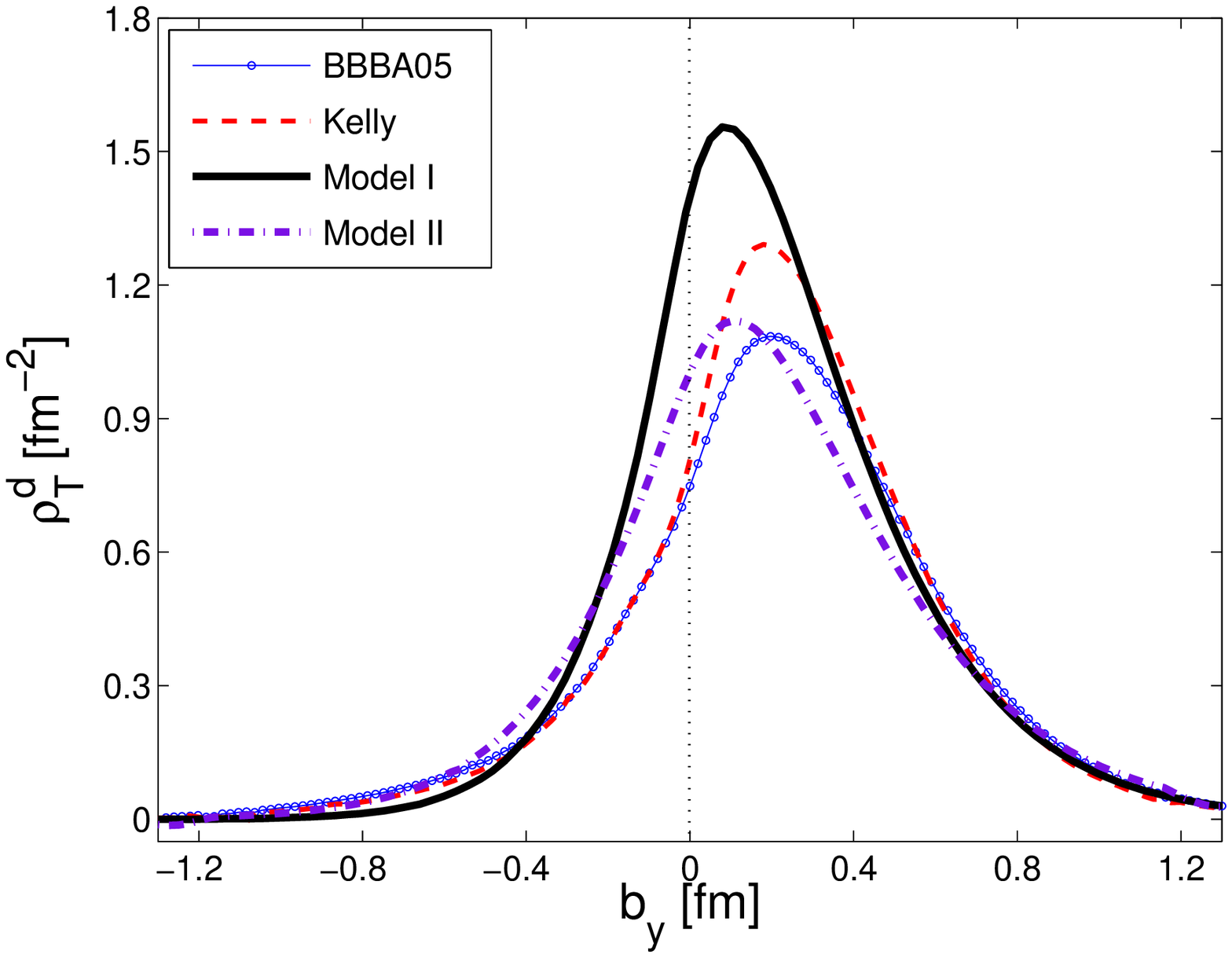}
\end{minipage}
\caption{\label{flavor_T}(Color online) The charge  densities for the transversely polarized (a) proton (b) neutron and (c) up (d) down quark charge  densities for the transversely polarized nucleon. 
Dashed line represents the parametrization of  Kelly \cite{kelly04},  and  line with circles represents the parametrization of  Bradford $at~el$ \cite{brad}; 
 the solid line is for Model-I and dot-dashed line is for Model-II.  }
\end{figure}
\begin{figure}[htbp]
\begin{minipage}[c]{0.98\textwidth}
\small{(a)}
\includegraphics[width=7cm,height=6cm,clip]{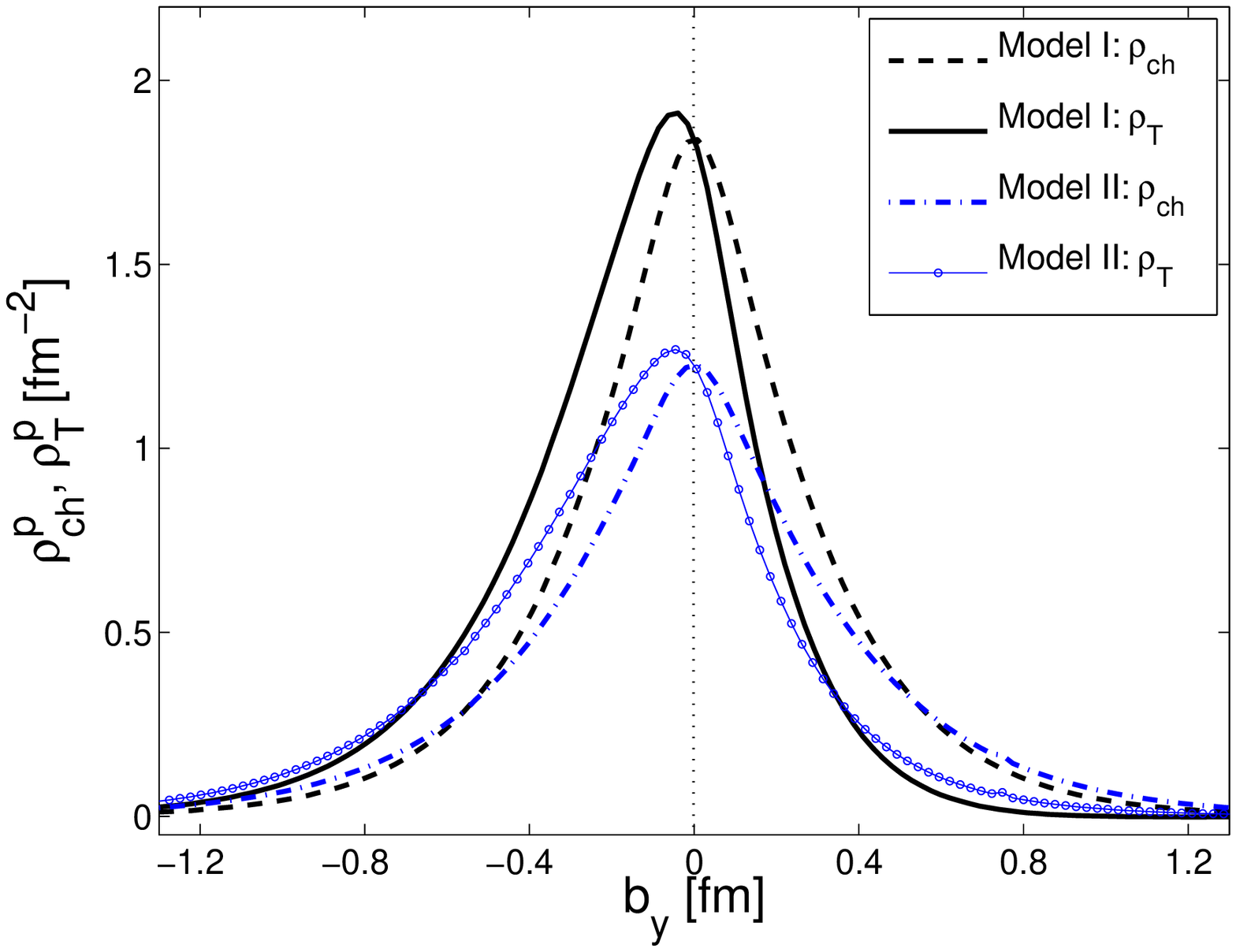}
\hspace{0.1cm}%
\small{(b)}\includegraphics[width=7cm,height=6cm,clip]{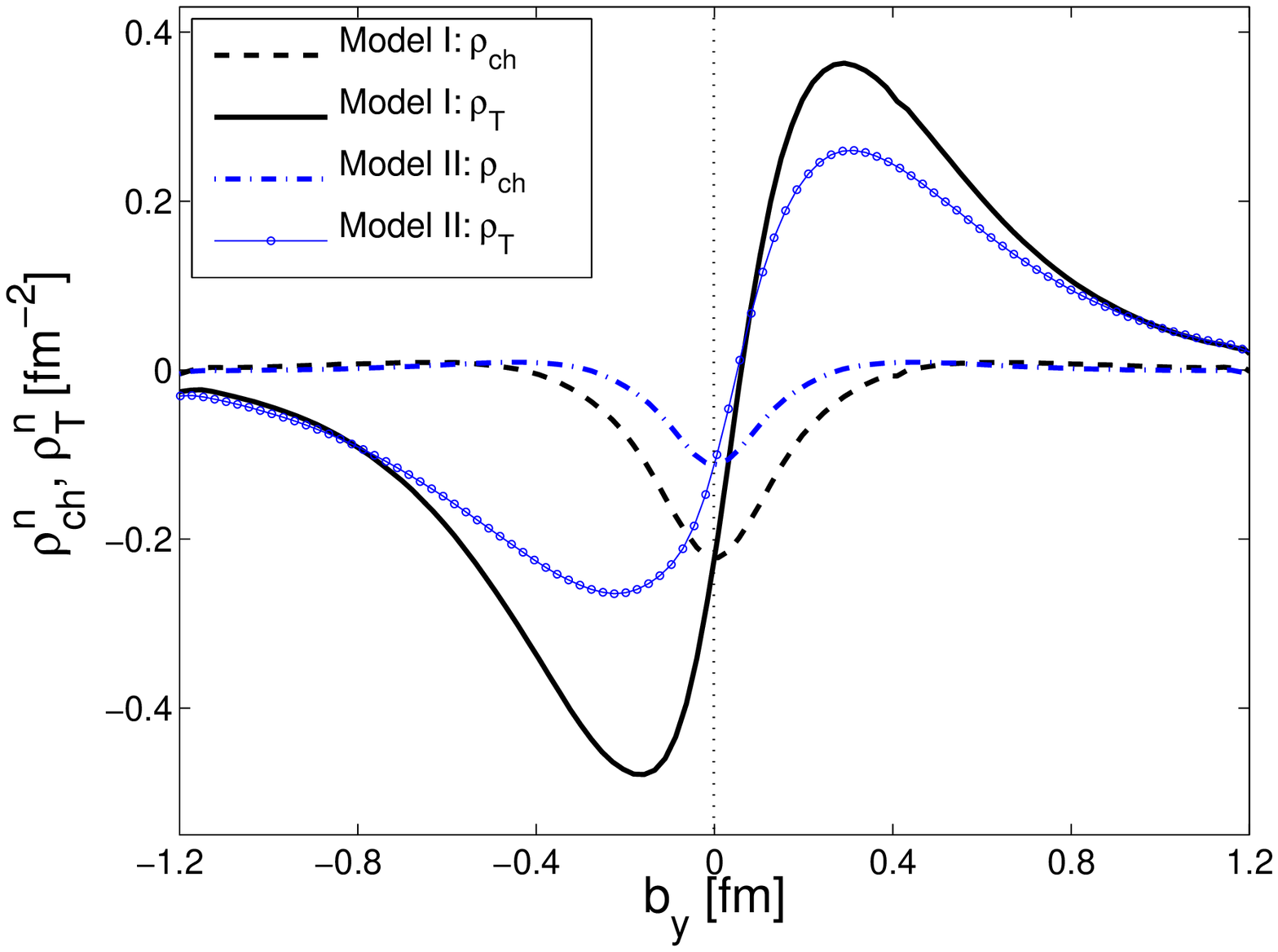}
\end{minipage}
\begin{minipage}[c]{0.98\textwidth}
\small{(c)}
\includegraphics[width=7cm,height=6cm,clip]{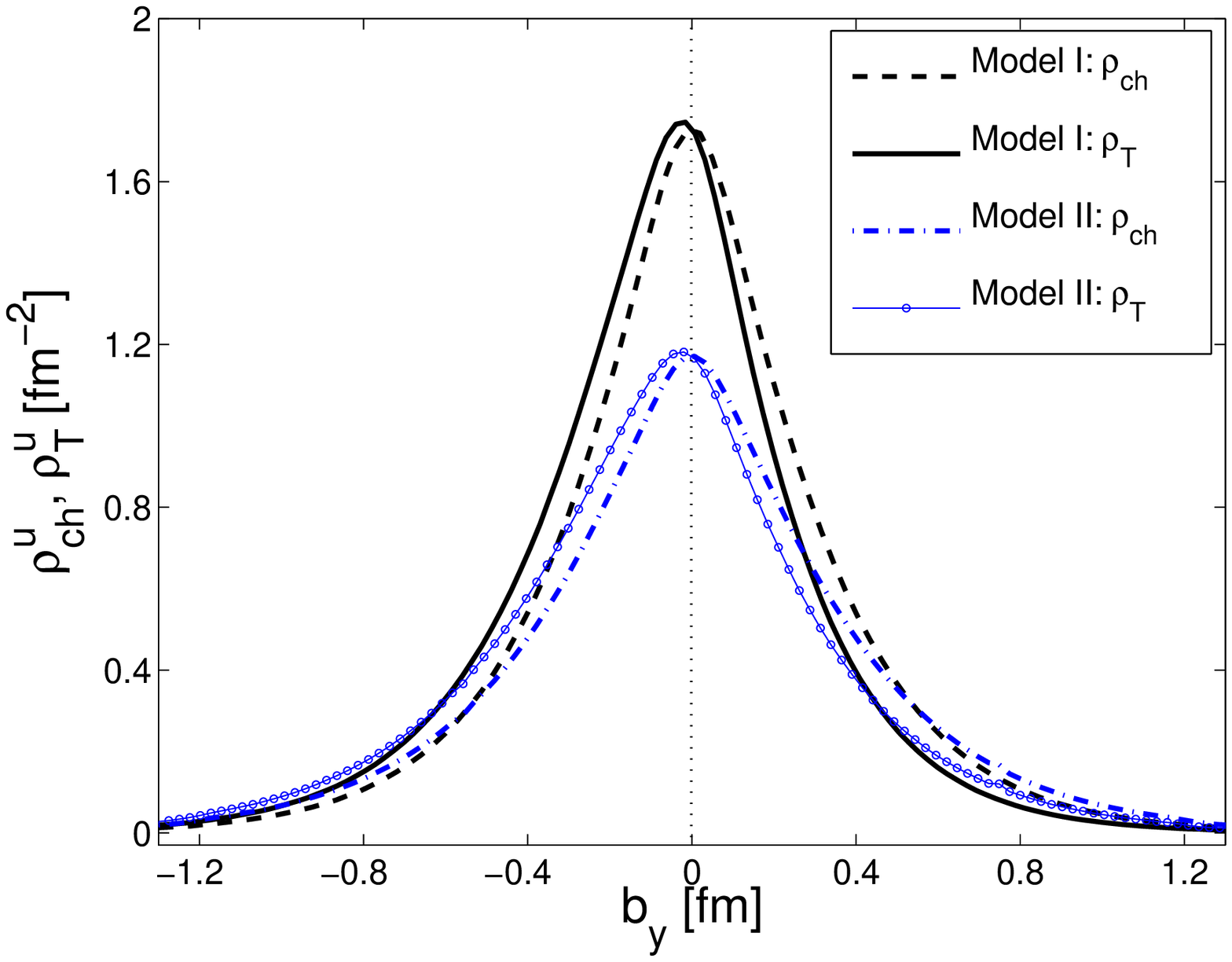}
\hspace{0.1cm}%
\small{(d)}\includegraphics[width=7cm,height=6cm,clip]{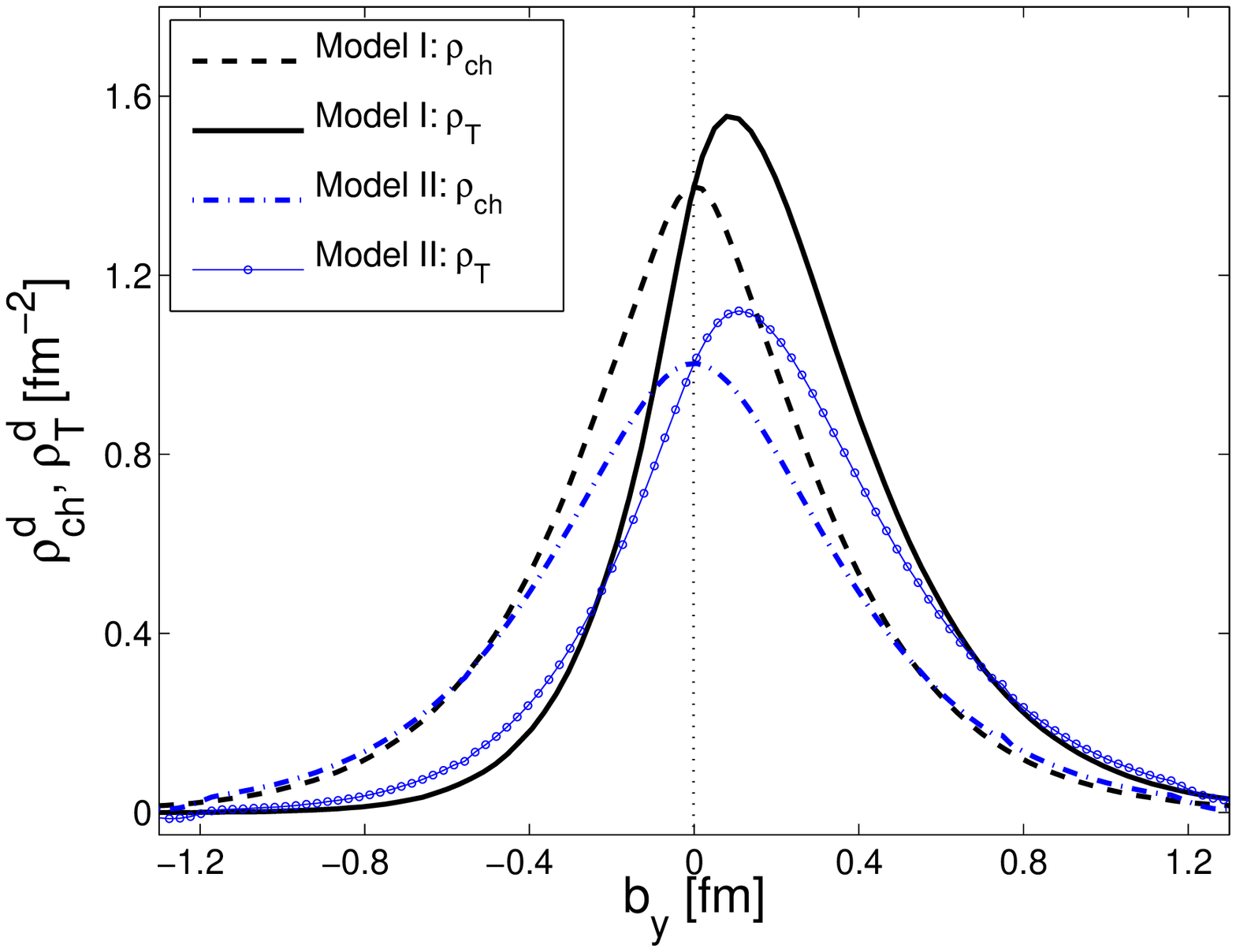}
\end{minipage}
\caption{\label{flavor_TADS}(Color online) The charge  densities for the transversely polarized (a) proton (b) neutron and (c) up (d) down quark charge  densities for the transversely polarized nucleon. The unpolarized charge densities are shown by the dashed line.}
\end{figure} 
\section{TRANSVERSE CHARGE AND MAGNETIZATION DENSITIES}\label{density}
The transverse charge density inside the nucleons is given by
\be
\rho_{ch}(b)
&=&\int \frac{d^2q_{\perp}}{(2\pi)^2}F_1(q^2)e^{iq_{\perp}.b_{\perp}}\nonumber\\
&=&\int_0^\infty \frac{dQ}{2\pi}QJ_0(Qb)F_1(Q^2),
\ee
where $b$ represents the impact parameter and $J_0$ is the cylindrical Bessel function of order zero. 
Similar formula for charge density for flavor $\rho_{fch}^q (b)$ can be written with $F_1$ is replaced
by $F_1^q$ . One can define the magnetization density in the similar fashion to have the formula
\be
\widetilde{\rho}_{M}(b) &= & \int \frac{d^2q_{\perp}}{(2\pi)^2}F_2(q^2)e^{iq_{\perp}.b_{\perp}}\nonumber\\
&=&\int_0^\infty \frac{dQ}{2\pi}QJ_0(Qb)F_2(Q^2),
\ee
Whereas,
\be
\rho_m(b)= -b\frac{\partial \widetilde{\rho}_M(b)}{\partial b}
=b\int_0^\infty \frac{dQ}{2\pi}Q^2J_1(Qb)F_2(Q^2),
\ee
has the interpretation of anomalous magnetization density \cite{miller10}. Since these quantities are not
directly measured in experiments, actual experimental data are not available. In \cite{venkat}, an
approximate estimation of the proton charge and magnetization densities has been done from
experimental form factor data. To get an insight into the contributions of the different
quark flavors, we evaluate the charge and anomalous magnetization densities for the up and
down quarks. 

We can define the decompositions of the transverse charge and magnetization densities for nucleons in the similar way as electromagnetic form factors \cite{Cates}. The charge densities decompositions in terms of two flavors can be written as
\be
 \rho_{ch}^p&=& e_u \rho_{fch}^u+e_d \rho_{fch}^d,\nonumber\\\label{ch_mag1}
 \rho_{ch}^n&=& e_u \rho_{fch}^d+e_d \rho_{fch}^u,
 \ee
where $e_u$ and $e_d$ are charge of $u$ and $d$ quarks respectively. We should remember that due to the charge and isospin symmetry, the $u$, $d$ quark densities in the proton are the same as the $d$, $u$ densities in the neutron as shown in \cite{miller07}
\be
 \rho_{ch}^u(b)&=&  \rho_{ch}^p+ \frac{\rho_{ch}^n}{2}=\frac{ \rho_{fch}^u}{2},\nonumber\\\label{ch_mag2}
 \rho_{ch}^d(b)&=&  \rho_{ch}^p+2 \rho_{ch}^n= \rho_{fch}^d,
 \ee
where $\rho_{ch}^q(b)$ is the charge density of each quark and $\rho^q_{fch}$ is the charge density for each flavor. We can also do the similar decompositions as Eq.(\ref{ch_mag1}) and Eq.(\ref{ch_mag2}) for $\rho_m$. 

We are not aware of any experimental data on transverse densities. Kelly\cite{kelly04} and Bradford {\it et al}\cite{brad} proposed two different phenomenological parameterizations of the nucleon form factor data. Here we calculate the transverse charge and magnetization densities from  these two parameterizations and compare with AdS/QCD predictions. Miller\cite{miller07} also used these parameterizations to evaluate the transverse charge densities of the nucleons.
In Fig.\ref{proton_flavors}(a) and (b) we show the charge and anomalous magnetization densities for proton.  The plots suggest that the Model-I agrees with the phenomenological parametrizations much better than the Model-II. The flavors contributions coming to proton densities from $e_{u/d}\rho_{fch}^{u/d}$ and $e_{u/d}{\rho}_{fm}^{u/d}$ are shown in Fig.\ref{proton_flavors}(c) and (d). Similarly the charge and anomalous magnetization densities for neutron and the flavors contributions $e_{d/u}\rho_{fch}^{u/d} $ and $e_{d/u}{\rho}_{fm}^{u/d}$ are shown in Fig.\ref{neutron_flavors}.  At small $b$, both the holographic models fail to reproduce the the neutron charge density.  Model-I  reproduces  the neutron magnetization density while model-II again fails to agree as shown in Fig.\ref{neutron_flavors}(b).
Model-I results for  the $u$ quark contributions  to the charge density  for both proton and neutron are in excellent agreement with the two different global parametrizations Kelly \cite{kelly04} and Bradford $at$ $el$ \cite{brad}. The $d$ quark contributions deviate form these two fits. It is not very surprising as it has been  already shown\cite{CM2} for Model-I that  the Dirac form factor for $d$ quark itself does not agree well with the experiment results. In case of anomalous magnetization both the quarks contributions  in proton and neutron agree quite well  with the fits.  The charge density for neutron (Fig.\ref{neutron_flavors}(a)) shows  a negatively charged core surrounded by a  ring of positive charge density (note that $b=0$ corresponds to the centre of the nucleon). In proton charge density the contribution of up quark is large enough compare to down quark but for neutron both contributions from $u$ and $d$ quark are comparable. For anomalous magnetization density of neutron, the $d$ quark 
contribution is quite high compared to $u$ quark. 
In Fig. \ref{flavors} (a) and (b) we show the individual quark's charge and anomalous magnetization densities. The charge density for  $d$ quark in Model-I  deviates form the fits but is in excellent agreement for $u$ quark. But again, as said before, deviation for $d$-quark in Model-I is expected. The anomalous magnetization densities in both $u$ and $d$ quarks in Model-I match very well with the fits. It is positive for $u$ quark but negative and larger for $d$ quark.  Model-II result for anomalous magnetization density of $d$-quark does not match so well with the phenomenological fits as Model-I.

For transversely polarized nucleon, the charge density is given by\cite{vande}
\be
\rho_T(b)=\rho_{ch}-\sin(\phi_b-\phi_s)\frac{1}{2M b}\rho_m\label{trans_pol},
\ee
where $M$ is the mass of nucleon and the transverse polarization of the nucleon is given by
 $S_\perp=(\cos\phi_s \hat{x}+\sin\phi_s\hat{y})$ and the transverse impact parameter $b_\perp=b(\cos\phi_b \hat{x} +\sin\phi_b\hat{y})$.  Without loss of generality, the polarization of the nucleon is taken along $x$-axis ie., $\phi_s=0$. The second term in Eq.(\ref{trans_pol}), provides the deviation from circular symmetry of the unpolarized charge density\cite{vande}. We show the charge densities for the transversely polarized proton and neutron in Fig.\ref{flavor_T}(a) and \ref{flavor_T}(b). The $u$ and $d$ quark charge  densities for the transversely polarized nucleon are shown in Fig.\ref{flavor_T}(c) and \ref{flavor_T}(d). Again, in Model-I, the densities for proton and $u$ quark are in  good agreement with the global parametrizations but deviate for neutron and $d$ quark.  Only for $d$-quark charge density as shown in Fig.\ref{flavor_T}(d), Model-II agrees with the phenomenological parametrizations better than Model-I. The comparison of charge densities for the transversely polarized and unpolarized 
proton is shown in Fig.\ref{flavor_TADS}(a) and the similar plot for neutron is shown in Fig.\ref{flavor_TADS}(b).  For the nucleons polarized along the $+x$ direction,, the 
charge densities are shifted towards negative $b_y$ direction. The deviation is much larger for the neutron compared to the proton. The behaviors are in agreement with  the results reported in \cite{vande,miller10,silva}. 
\begin{figure}[htbp]
\begin{minipage}[c]{0.98\textwidth}
\small{(a)}
\includegraphics[width=7cm,height=6cm,clip]{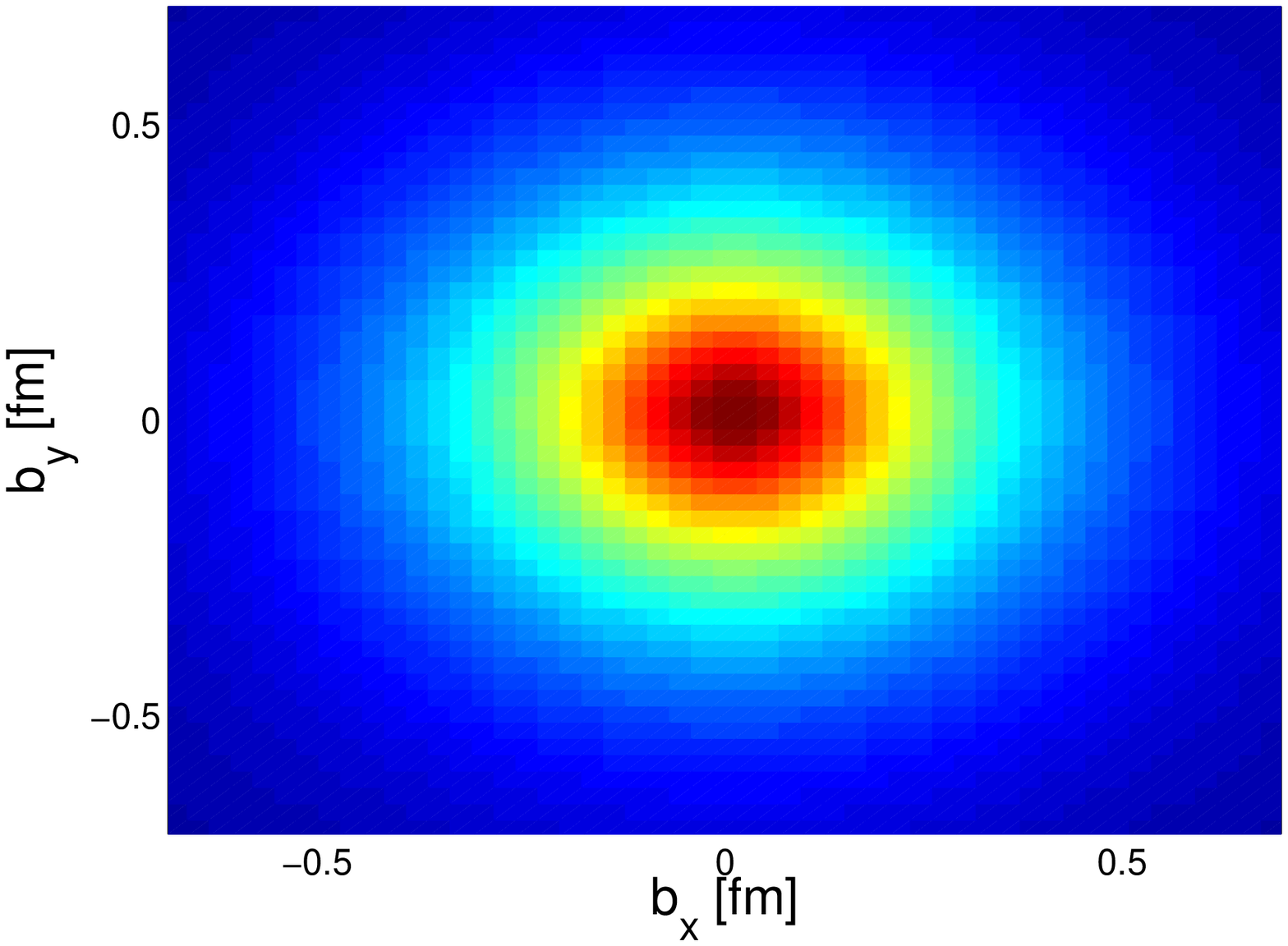}
\hspace{0.1cm}%
\small{(b)}\includegraphics[width=7cm,height=6cm,clip]{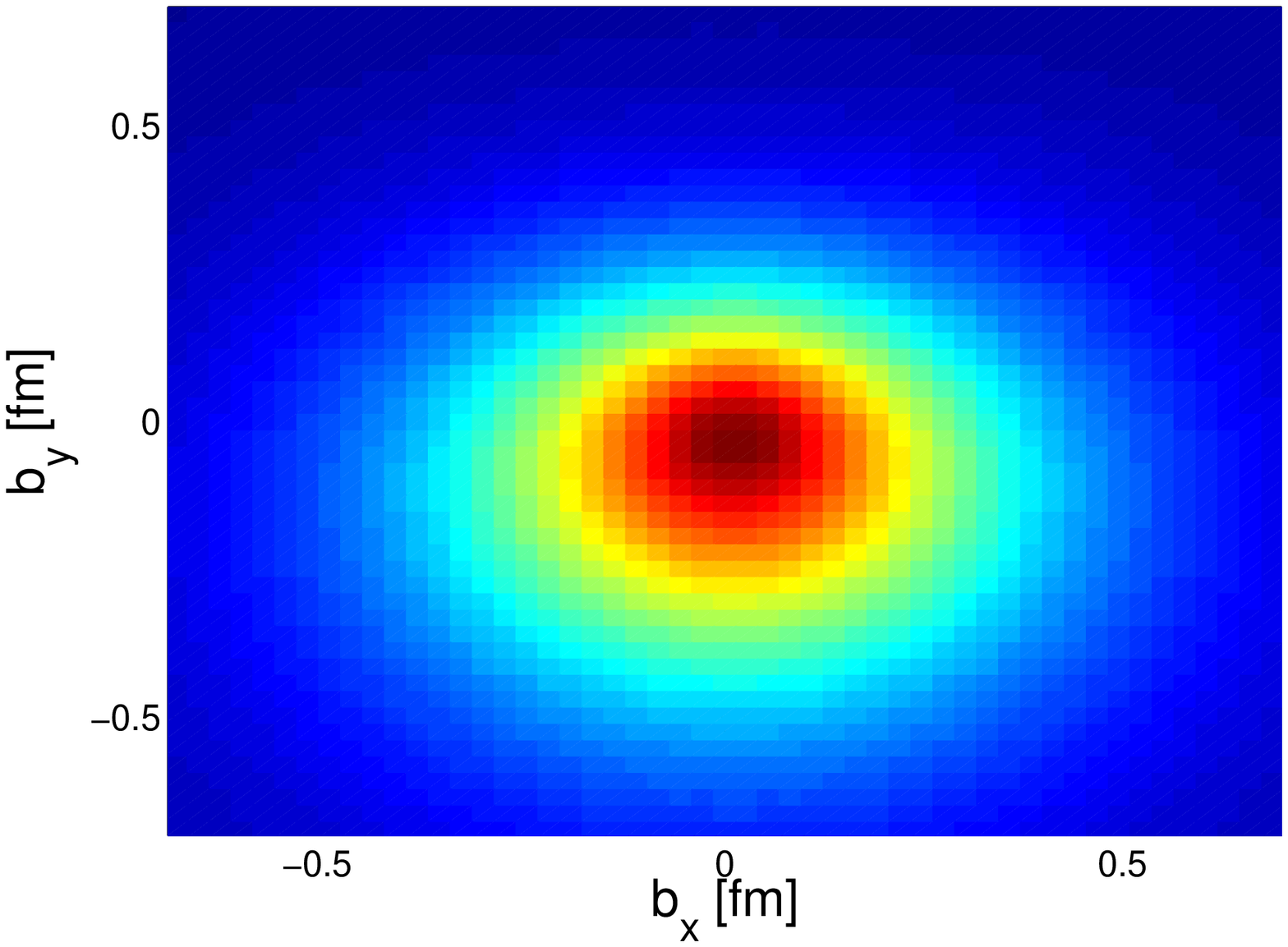}
\end{minipage}
\begin{minipage}[c]{0.98\textwidth}
\small{(c)}
\includegraphics[width=7cm,height=6cm,clip]{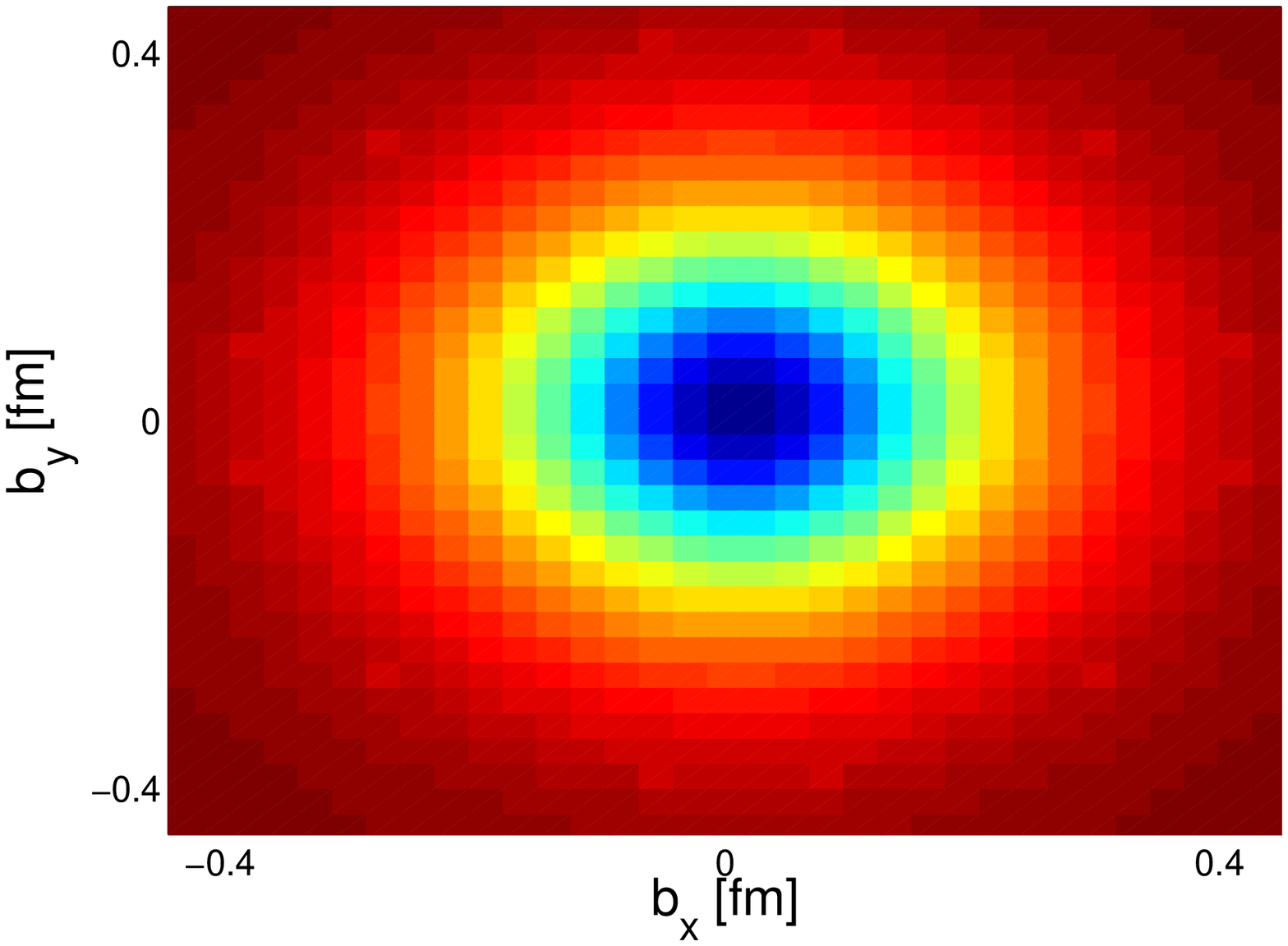}
\hspace{0.1cm}%
\small{(d)}\includegraphics[width=7cm,height=6cm,clip]{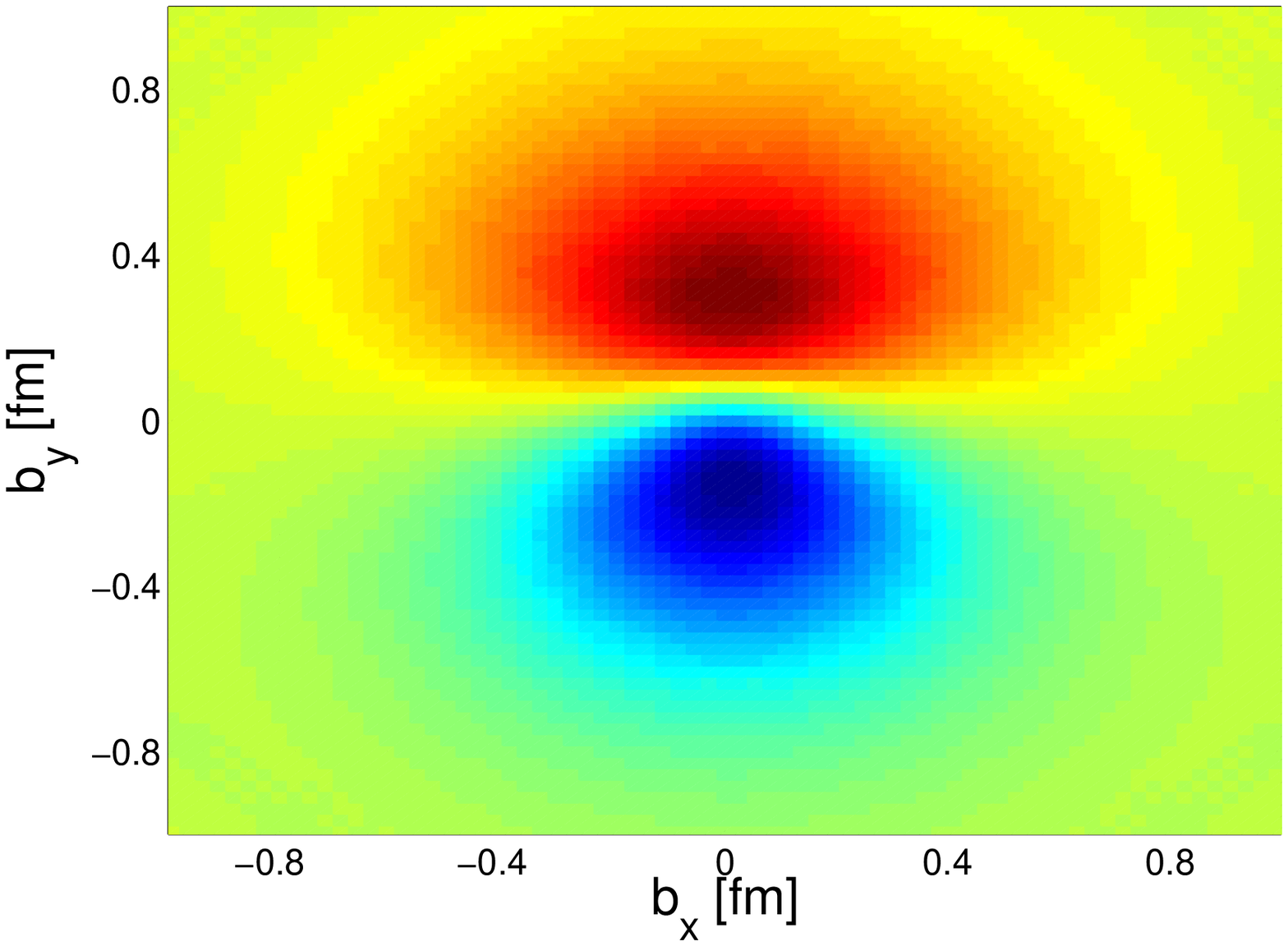}
\end{minipage}
\caption{\label{nucleons_top}(Color online) The charge  densities in the transverse plane for  the (a) unpolarized  proton (b) transversely polarized  proton and  (c) unpolarized neutron and  (d) transversely polarized neutron. Transverse polarization is along $x$ direction. }
\end{figure} 
\begin{figure}[htbp]
\begin{minipage}[c]{0.98\textwidth}
\small{(a)}
\includegraphics[width=7cm,height=6cm,clip]{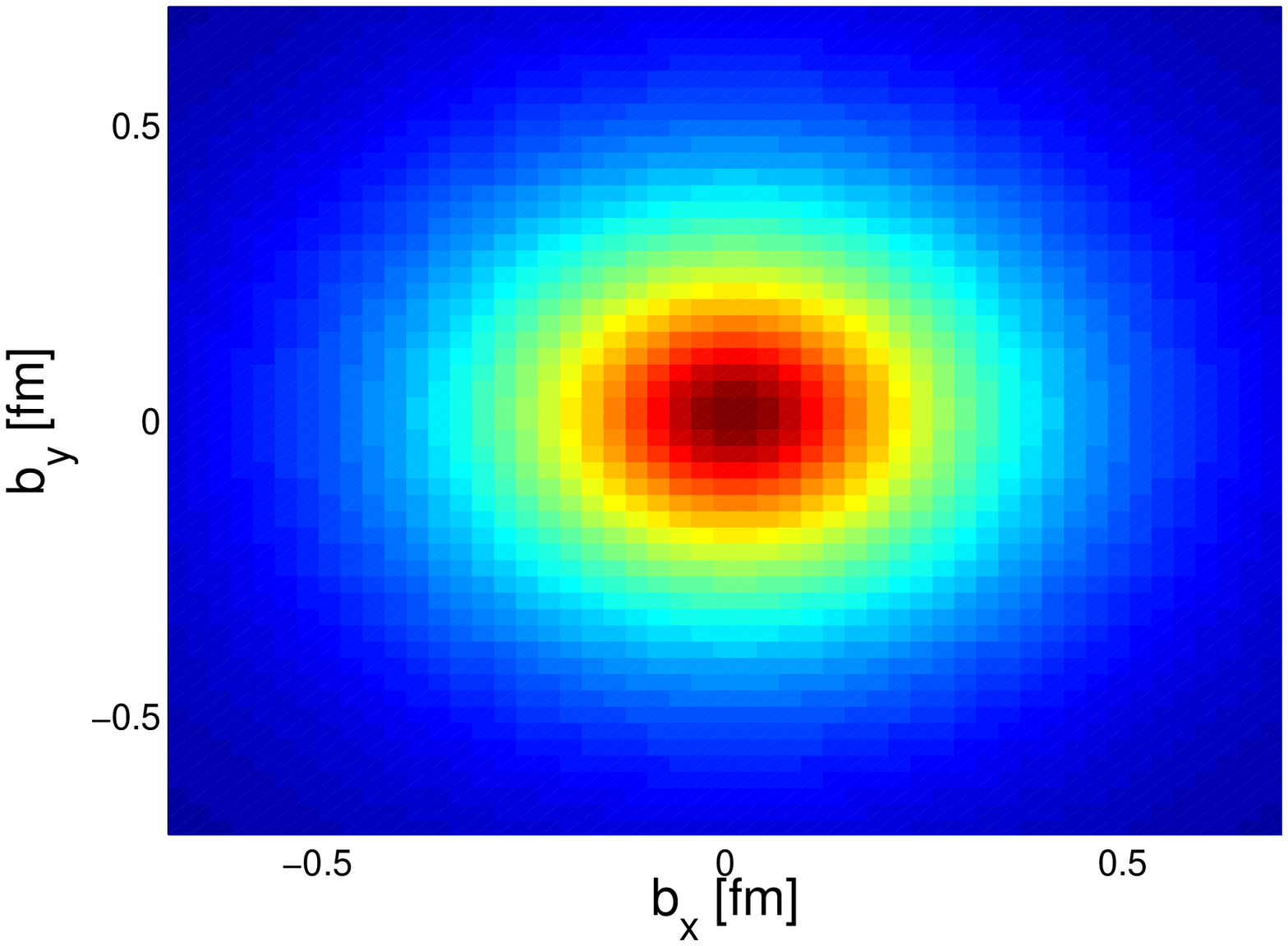}
\hspace{0.1cm}%
\small{(b)}\includegraphics[width=7cm,height=6cm,clip]{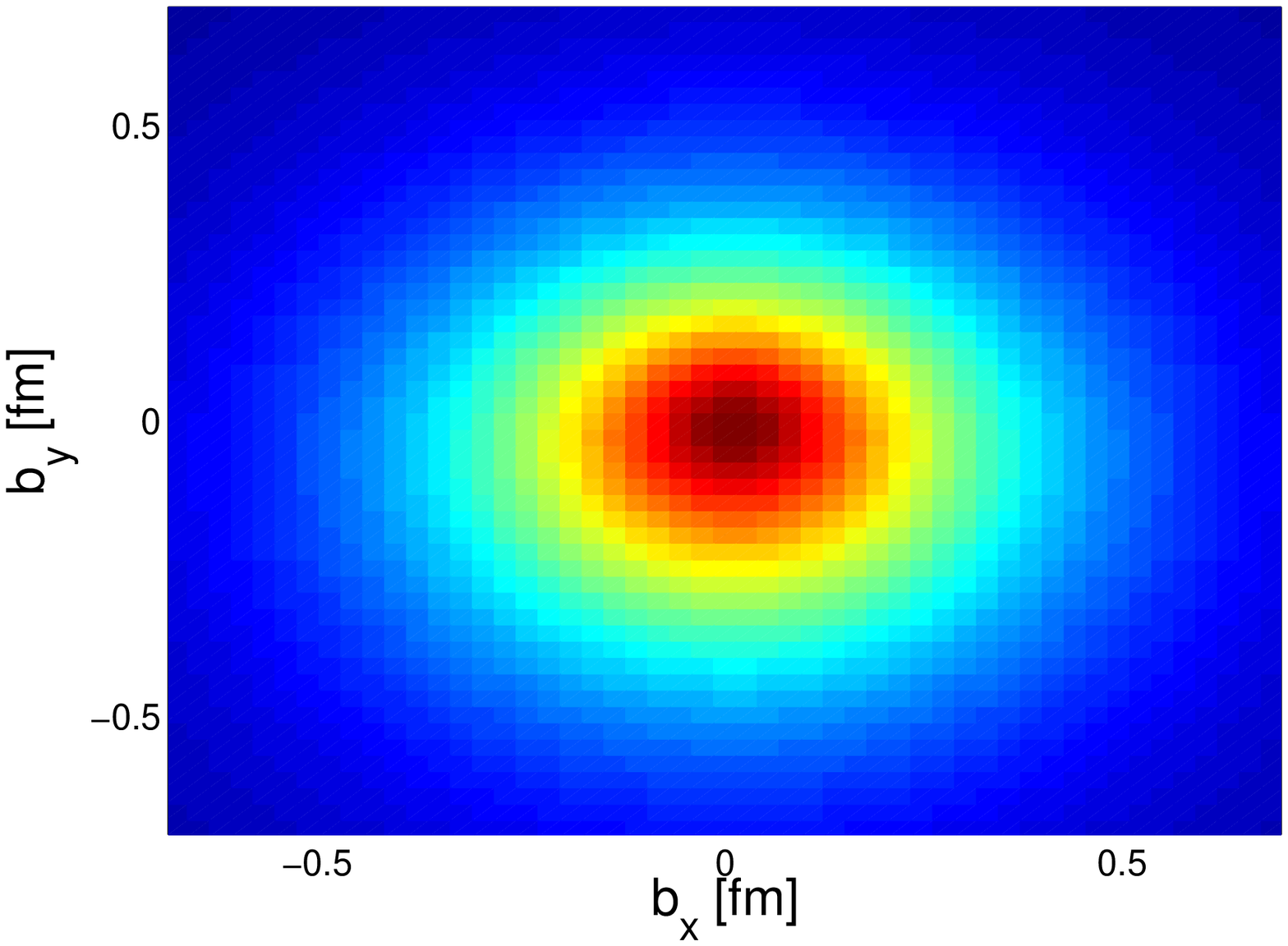}
\end{minipage}
\begin{minipage}[c]{0.98\textwidth}
\small{(c)}
\includegraphics[width=7cm,height=6cm,clip]{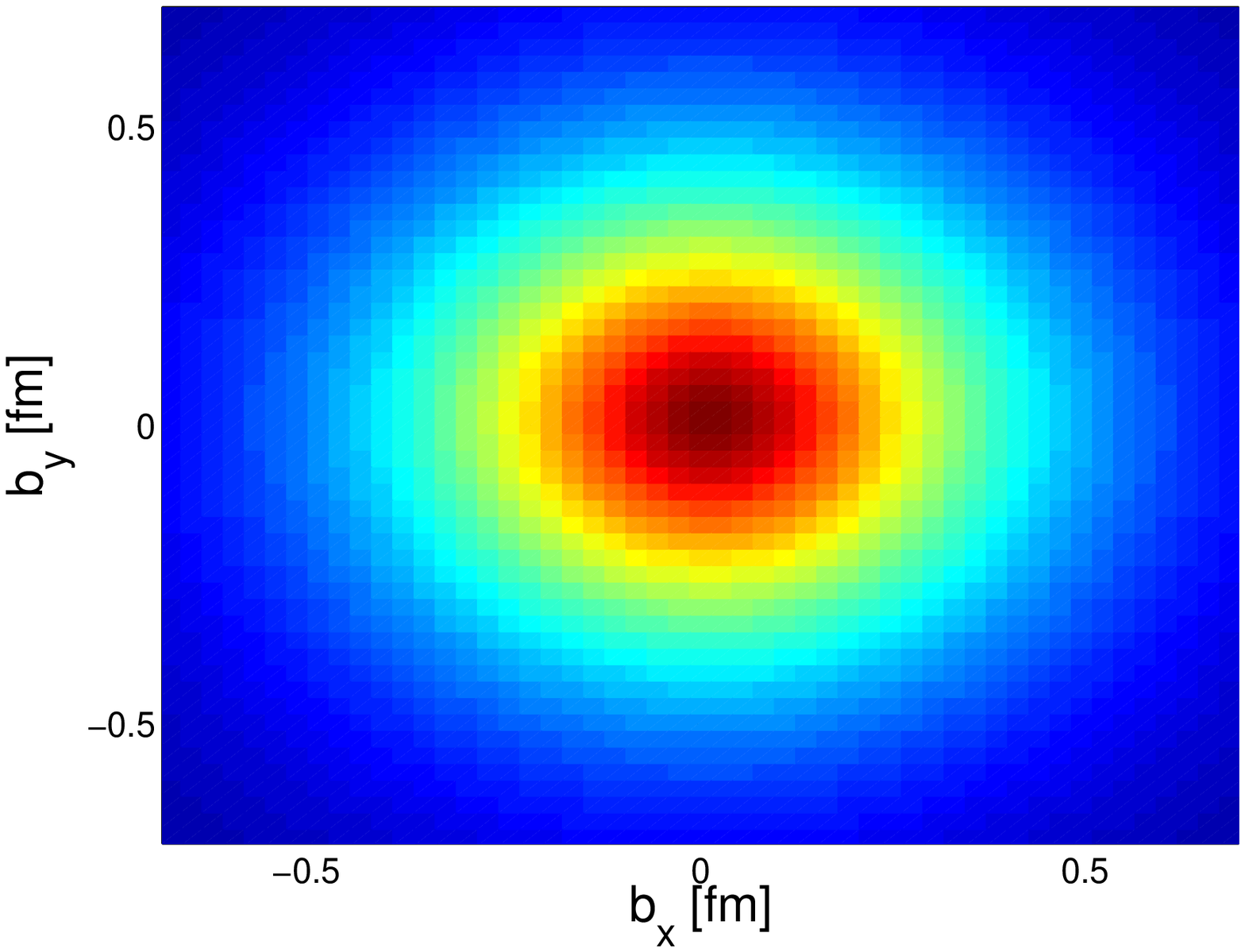}
\hspace{0.1cm}%
\small{(d)}\includegraphics[width=7cm,height=6cm,clip]{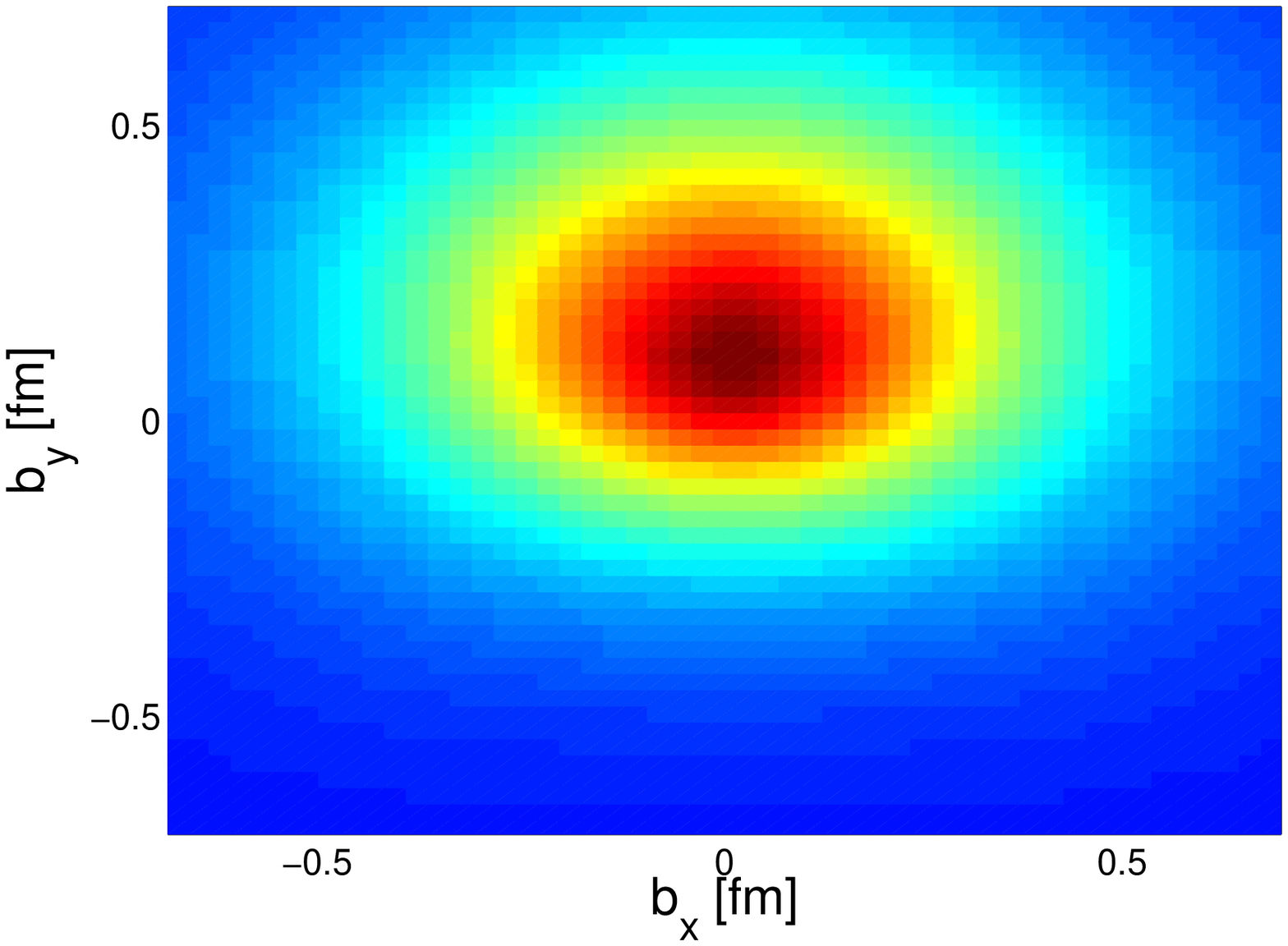}
\end{minipage}
\caption{\label{flavor_top}(Color online) The charge  densities in the transverse plane of $u$ quark for (a) unpolarized (b) transversely polarized nucleon and $d$ quark for (c) unpolarized (d) transversely polarized nucleon. Polarization is along $x$-direction.}
\end{figure} 
We compare the up and down quark charge densities for the transversely polarized and unpolarized nucleon
in Fig.\ref{flavor_TADS}(c) and \ref{flavor_TADS}(d). The deviation or distortion from the symmetric unpolarized density is more for down quark than the up quark. The shifting of charge density for the nucleons polarized in $+x$ direction, is towards positive $b_y$ direction for down quark but opposite for up quark. In Fig. \ref{nucleons_top}, we have shown a top view plot of charge densities in the transverse plane for (a) unpolarized  proton and (b) proton polarized along $x$-direction in model-I. Similar plots for neutron are shown in Fig. \ref{nucleons_top} (c) and (d).  Due to large anomalous magnetic moment which produces an induced electric dipole moment in $y$-direction, the distortion is more in the case of neutron\cite{vande}.
The top view plots for the  $u$ and $d$ quarks charge densities in the transverse plane for both unpolarized and transversely polarized nucleon are shown in Fig.{\ref{flavor_top}(a)-(d).
 
\vskip0.2in
\noindent
\section{\bf Summary}\label{concl}
In this paper, we have presented a detailed study and comparison  of the charge and anomalous magnetization densities for nucleons in the transverse plane in two models  in AdS/QCD. We  have also compared our results with the two standard phenomenological parametrizations  of the form factors. 
Both the unpolarized and transversely polarized nucleons have been considered in this work. The unpolarized densities are symmetric in transverse plane while for the transversely polarized nucleons they become distorted. If the nucleon is polarized along $x$ direction, the densities get shifted towards negative $y$-direction. We have also studied the flavor decompositions of the transverse densities i.e., the charge and anomalous magnetization densities for individual  $u$ and $d$ quark flavors.  Our analysis shows that Model-I reproduces the data much better than the Model-II. The agreement is not so good for $d$ quark which is consistent with the findings in   \cite{CM2},  where the form factors for the $d$ -quark were shown to deviate from the experimental results.
For transversely polarized nucleon, the distortion in $d$ quark charge density is found to be stronger than that for $u$ quark and shifted in opposite direction  to each other.

Acknowledgement:  We thank V.E. Lyubovitskij, S. J. Brodsky and G. F. de Teramond for insightful correspondence on formfactors in AdS/QCD.}




\end{document}